\documentclass[journal]{IEEEtran}
\usepackage{microtype} % Fine font adjustments
\usepackage{siunitx}
\usepackage{multirow}
\usepackage{url}

\usepackage{cite}
\ifCLASSINFOpdf
   \usepackage[pdftex]{graphicx}
   \graphicspath{{./pdf/}{./figures/}}
   \DeclareGraphicsExtensions{.pdf,.jpeg,.png, .tif}
\else
\fi
\usepackage[cmex10]{amsmath}
\usepackage{array}
\usepackage{fixltx2e}

\ifCLASSOPTIONcompsoc
  \usepackage[caption=false,font=normalsize,labelfont=sf,textfont=sf]{subfig}
\else
  \usepackage[caption=false,font=footnotesize]{subfig}
\fi

\usepackage{graphicx}
\usepackage[linesnumbered,ruled,vlined]{algorithm2e}
\usepackage{algorithmicx}
\usepackage{algpseudocode}
\usepackage{amsfonts,amssymb,amsmath}
\usepackage{bm,bbm}
\usepackage{subfig}
\usepackage{array}
\usepackage{booktabs}
\usepackage[switch]{lineno}

\DeclareMathOperator*{\argmax}{\arg\!\max}

\usepackage{epstopdf}
\DeclareMathAlphabet{\mathbfit}{OML}{cmm}{b}{it}

\graphicspath{{./Figures_re/}}

\begin{document}

\title{Modifying the Yamaguchi Four-Component Decomposition Scattering Powers Using a Stochastic Distance}

\author{Avik~Bhattacharya,~\IEEEmembership{Member,~IEEE,}
        Arnab~Muhuri,
        Shaunak~De, Surendar~Manickam, and\\
        Alejandro C.~Frery,~\IEEEmembership{Senior Member,~IEEE}% <-this % stops a space
\thanks{A.~Bhattacharya, A.~Muhuri, S.~De and M.~Surendar are with the Center 
of Studies in Resources Engineering, Indian Institute of Technology Bombay, Mumbai, MH, 400076 India \mbox{e-mail:~avikb@csre.iitb.ac.in}.}% <-this % stops a space
\thanks{ A.C. Frery is with the Universidade Federal de Alagoas, Macei\'o, Brazil}% <-this % stops a space 
%\thanks{Manuscript received April 19, 2005; revised December 27, 2012.}
%\thanks{ }
}

\maketitle

\begin{abstract}
Model-based decompositions have gained considerable attention after the initial work of Freeman and Durden. This decomposition which assumes the target to be reflection symmetric was later relaxed in the Yamaguchi $\emph{et al.}$ decomposition with the addition of the helix parameter. Since then many decomposition have been proposed where either the scattering model was modified to fit the data or the coherency matrix representing the second order statistics of the full polarimetric data is rotated to fit the scattering model. In this paper we propose to modify the Yamaguchi four-component decomposition (Y4O) scattering powers using the concept of statistical information theory for matrices. In order to achieve this modification we propose a method to estimate the polarization orientation angle (OA) from full-polarimetric SAR images using the Hellinger distance. In this method, the OA is estimated by maximizing the Hellinger distance between the un-rotated and the rotated $T_{33}$ and the $T_{22}$ components of the coherency matrix $\mathbf{[T]}$. Then, the powers of the Yamaguchi four-component model-based decomposition (Y4O) are modified using the maximum relative stochastic distance between the $T_{33}$ and the $T_{22}$ components of the coherency matrix at the estimated OA. The results show that the overall double-bounce powers over rotated urban areas have significantly improved with the reduction of volume powers. The percentage of pixels with negative powers have also decreased from the Y4O decomposition. The proposed method is both qualitatively and quantitatively compared with the results obtained from the Y4O and the Y4R decompositions for a Radarsat-2 C-band San-Francisco dataset and an UAVSAR L-band Hayward dataset.
\end{abstract}
 

% \fi
\IEEEpeerreviewmaketitle
%\boldmath

\section{Introduction}
\PARstart{P}{olarimetric}
target decomposition is a technique to characterize scattering mechanisms from polarimetric synthetic aperture radar (SAR) data. 
Target decomposition techniques can be broadly classified into two categories: 
(1)~coherent decomposition techniques which utilize the information contained in a target scattering matrix $\mathbf{[S]}$, and 
(2)~incoherent decomposition techniques which utilize the second-order statistics in terms of covariance matrices ($\mathbf{[C]}$ or $\mathbf{[T]}$) derived from the scattering matrix $\mathbf{[S]}$. 
Furthermore, incoherent decomposition techniques can be subdivided into eigenvalue/eigenvector based, and model-based. 
Eigenvalue/eigenvector based decompositions provide a unique solution in terms of the scattering mechanisms~\cite{CLOUDE97,TOUZI2007}. 
However, Ref.~\cite{Alvarez-Perez2011} raised questions regarding the assignment of each eigenvector to one of the independent scattering mechanisms by the spectral decomposition of the coherency matrix. 
Nonetheless, the solutions provided by model-based decompositions depend on the assumptions made about the physical scattering model.  

Model-based decompositions have gained considerable attention after the initial work of Freeman and Durden~\cite{freeman98}. 
Their decomposition assumes the target to be reflection symmetric, i.e., that the co-polarized and the cross-polarized components are always uncorrelated. 
This assumption was later relaxed in the Yamaguchi~\emph{et al.} decomposition (Y4O)~\cite{Yamaguchi05} which included the helical scattering as a fourth component. 
These two decompositions are widely used in both practice and literature because of their simplicity and computational ease. 
Besides these, the decomposition of Arii~\emph{et al.}~\cite{arii2010general,Arii11} and Neumann~\emph{et al.}~\cite{neumann2010estimation,neumann2010extraction} introduced different scattering models for vegetation volume component. 
Jagdhuber~\emph{et al.}~\cite{Jagdhuber2013} proposed a full-polarimetric decomposition with multi-angular data acquisition to estimate soil moisture from bare ground and vegetated soils. 
Cui~\emph{et al.}~\cite{Cui2014} proposed a complete and a exact decomposition of the coherency matrix into a volume and two single scattering with non-negative scattering powers. 
Lee~\emph{et al.}~\cite{Lee2014} investigated the shortcomings of model-based decompositions, and suggested several models to alleviate them. 
Recently, Jagdhuber~\emph{et al.}~\cite{Jagdhuber2014} developed a hybrid model-based and eigenvalue/eigenvector-based polarimetric decomposition technique with generalized volume model for soil moisture estimation under vegetation cover. 

A major advancement was made with the orientation compensation application in model-based decompositions. 
This was necessary because of the fact that a target with different orientations in the plane orthogonal to the radar line of sight (LOS) will have different polarimetric responses. 
A number of decomposition methods with orientation compensation have been proposed~\cite{Lee2011,An10,YAMAGUCHI2011,singh13,Arii11,Chen14}. 
The fundamental idea behind such compensation is to minimize the cross-polarization component. 
The orientation compensation can, to a certain extent, reduce the overestimation of the volume power in a model-based decomposition and increase the double-bounce power. 
Most notably among these methods are Lee~\emph{et al.}~\cite{Lee2011}, the three-component model-based decomposition by An~\emph{et al.}~\cite{An10}, the orientation compensated four-component decomposition by Yamaguchi~\emph{et al.} (Y4R)~\cite{YAMAGUCHI2011}, and the generalized four-component decomposition by real and complex rotation of the coherency matrix by Singh~\emph{et al.}~\cite{singh13}. 
The decomposition methods of Arii~\emph{et al.}~\cite{Arii11} and Chen~\emph{et al.}~\cite{Chen14} also use orientation angle in their scattering models.

In general, orientation angle (OA) estimation methods can be broadly categorized into two groups according to the input data:
(1)~OA derived from a Digital Elevation Model (DEM), and 
(2)~OA derived from PolSAR data.  
The slope and the azimuth obtained from the DEM is used to estimate the OA. 
Apart from the DEM derived orientation angle there are few other methods available in the literature which directly use PolSAR data to compute the orientation angle. 
The phase difference between the RR-LL (Right-Right and Left-Left) circular polarizations has been used in~\cite{Lee2000} to estimate the orientation angle. 
The polarization orientation shift is used to infer terrain slopes~\cite{Schuler96}. 
In a later study, Kimura~\emph{et al.}~\cite{Kimura08} computed the shifts in the polarization orientation angle in built-up areas. 
The beta ($\beta$) angle obtained from the eigen-decomposition of the coherency matrix in Cloude-Pottier decomposition~\cite{CLOUDE97} is also used as a measure of the orientation angle. 
For deterministic (coherent) scatterers, the orientation angle (diagonalization angle) is obtained from the Cameron~\emph{et al.} decomposition~\cite{Cameron96}. 
Xu~\emph{et al.}~\cite{Xu05} estimate the target deorientation angle by minimizing the cross-polarization.

In this paper we propose to modify the Yamaguchi four-component decomposition (Y4O) scattering powers using the concept of statistical information theory for matrices. 
In order to achieve this modification, we first introduce the concept of stochastic divergence in Section~\ref{Sec:StochasticDistances}. 
In Section~\ref{Sec:OAEstimation}, we propose a method to estimate the polarization OA from full-polarimetric SAR images using a stochastic distance. 
In this method, the OA is estimated by maximizing the stochastic distance between the un-rotated and the rotated $T_{33}$ and the $T_{22}$ elements of the coherency matrix $\mathbf{[T]}$. 
In Section~\ref{Sec:sd-y4o}, the powers of the Yamaguchi four-component model-based decomposition (Y4O) are then modified using the relative stochastic distance between the $T_{33}$ and the $T_{22}$ elements of the coherency matrix at the estimated OA. 
The results show that the overall double-bounce powers over rotated urban areas improve significantly with the reduction of volume powers. 
The percentage of pixels with negative powers is also reduced from the Y4O decomposition. 
In Section~\ref{Sec:results_discussion}, the proposed method is both qualitatively and quantitatively compared with the results obtained from the Y4O and the Y4R decompositions. 

\section{Stochastic Distances}\label{Sec:StochasticDistances}

In this section we introduce our main tool: the concept of stochastic divergence~\cite{liese2006divergences}.
Such class of techniques has found many application ranging from classification~\cite{puig2003pixel}, cluster analysis~\cite{mak1996phone} and goodness-of-fit tests~\cite{zografos1990divergence}. 
Specifically in SAR image processing and analysis, these statistical separability measures have been used for 
estimation~\cite{ParameterEstimationSARStochasticDistancesKernels},
classification and segmentation~\cite{NascimentoCintraFreryIEEETGARS,EntropyBasedStatisticalAnalysisPolSAR,ClassificationPolSARSegmentsMinimizationWishartDistances},
noise reduction~\cite{TorresPolarimetricFilterPatternRecognition}, and 
change detection~\cite{APSAR2013ChangeDetection}. 

Measures of divergence among probability distributions are an intuitive approach for comparing models. 
In this work, we use the concept of stochastic divergence to estimate the polarization OA from full-polarimetric SAR images.

Goudail and R\'efr\'egier~\cite{GoudailRefreger:2004} used numerical integration to compute the Kullback and Bhattacharyya distances to derive a scalar measure of contrast between areas under the 2D circular Gaussian model.
Nascimento~\emph{et al.}~\cite{NascimentoCintraFreryIEEETGARS} used the same approach for intensity data under the $\mathcal G_I^0$ distribution, and compared the performance of test statistics derived from a number of ($h$-$\phi$) distances (Kullback-Leibler, R\'enyi, Hellinger, Bhattacharyya, Jensen-Shannon, Arithmetic-Geometric, Triangular, and Harmonic Mean) as features for image classification.
Cintra \emph{et al.}~\cite{ParametricNonparametricTestsSpeckledImagery} compared some of these measures with parametric and nonparametric tests, and they outperformed other techniques in terms of efficiency and robustness.

Frery~\emph{et al.}~\cite{EntropyBasedStatisticalAnalysisPolSAR,
AnalyticExpressionsStochasticDistancesBetweenRelaxedComplexWishartDistributions}
obtained analytic expressions for entropies and distances between Wishart models for full PolSAR data.
Besides being able to build adequate classificators, as the one presented by Silva~\emph{et al.}~\cite{ClassificationPolSARSegmentsMinimizationWishartDistances}, such expressions lead to efficient edge detectors~\cite{EdgeDetectionDistancesEntropiesJSTARS} and nonlocal means filters~\cite{TorresPolarimetricFilterPatternRecognition}.

The cornerstone of these results is the family of ($h$-$\phi$) divergences, as defined by Salicr\'u~\emph{et al.}~\cite{Salicru1994}.
Consider two probability distributions $\mathcal D_1$ and $\mathcal D_2$ defined on the same support $\mathcal S$ and characterized, without loss of generality, by the densities $f_1$ and $f_2$.
Given any strictly increasing (decreasing, respectively) function $h\colon \mathbbm R_+\to[0,\infty]$ such that $h(0)=0$, and any convex  (concave, resp.) function $\phi\colon\mathbbm R_+\to[0,\infty]$ then
\begin{equation}
D^h_\phi(\mathcal D_1,\mathcal D_2) = 
h \bigg(
	\int_{\mathcal S} 
		\phi \Big(
			\frac{f_1}{f_2}
		\Big) f_2
\bigg)
	\label{Eq:Dhphi}
\end{equation}
is the $D^h_\phi$ divergence between the distributions.
Mild properties are required for the defining functions $h$ and $\phi$, namely that $\lim_{x\to0^+} \phi(x)$ exists, that $0\phi(0/0)\equiv 0$, and that for any $a>0$ holds that $\lim_{\epsilon\to0^+} \epsilon \phi(a/\epsilon) = a \lim_{x\to\infty} \phi(x)/x = 0\phi(a/0)$.

Any $D^h_\phi$ divergence can be turned into a $d^h_\phi$ distance making $d^h_\phi(\mathcal D_1,\mathcal D_2) =\big(
D^h_\phi(\mathcal D_1,\mathcal D_2) + D^h_\phi(\mathcal D_2,\mathcal D_1)
\big)/2$.
These are stochastic distances as they have the properties of symmetry, non-negativity and identity of indiscernibles.
Besides that, as attested by the aforementioned literature, they can be turned into convenient tools for image analysis.

Adequate choices of $h$ and $\phi$ lead to many well known distances, being the Hellinger distance the one we will use in this work.
It is obtained with $h(x)=x/2$ and $\phi(x)=(x^{1/2}-1)^2$, 
then~\eqref{Eq:Dhphi} becomes
$$
d^h_\phi(\mathcal D_1,\mathcal D_2) = 1 - \int \sqrt{f_1 f_2},
$$
as it is also a distance.

In the following we will instantiate this distance under the most important model for ful PolSAR data: the Wishart law.

Denote $\mathbfit{z}=[S_{1},S_{2},\dots,S_{p}]^{T}$ a complex random vector with $p$ polarizations ($p=3$ for reciprocal medium or monostatic radar) where $S_{1}$, $S_{2}$ and $S_{3}$ are either $S_{HH}$, $\sqrt{2}S_{HV}$, $S_{VV}$ or $S_{HH}+S_{VV}$, $S_{HH}-S_{VV}$, $2S_{HV}$ are the elements of a scattering vector in the Lexicographic and Pauli bases, respectively. 
In PolSAR data analysis, $\mathbfit{z}$ is often assumed to obey a zero mean multivariate complex Gaussian distribution characterized by the following density: 
\begin{equation}
f(\mathbfit{z};\mathbf{\Sigma})=\frac{1}{\pi^{p}\left|\mathbf{\Sigma}\right|}\mbox{exp}{(-\mathbfit{z}^{*T}\mathbf{\Sigma}^{-1}\mathbfit{z})},
\label{eq:mul_gaussian}
\end{equation}
where $\mathbf{\Sigma}$ is the Hermitian positive definite covariance matrix, $v^*$ denotes the conjugate of $v$, and $v^T$ its transpose.
In order to increase the signal-to-noise ratio, $L$ independent and identically distributed samples are averaged to form the $L$-looks covariance matrix
\begin{equation}
\mathbfit{Z}=\frac{1}{L}\sum\limits_{\ell=1}^L \mathbfit{z}(\ell)\mathbfit{z}(\ell)^{*T}.
\label{eq:L_look_cm}
\end{equation} 

The $p\times p$ Hermitian positive definite matrix $\mathbfit{Z}$ follows a scaled complex Wishart distribution whose density is
\begin{equation}
f(\mathbfit{Z})=\frac{L^{pL}\left| \mathbfit{Z}\right|^{L-p} \exp[-L\mbox{Tr}(\mathbf{\Sigma}^{-1}\mathbfit{Z})]}{\Gamma_{p}(L)\left|\mathbf{\Sigma}\right|^L},
\label{eq:wishart}
\end{equation}
where $\mbox{Tr}$ is the trace and $\Gamma_{p}(L) = \pi^{\frac{p(p-1)}{2}}\prod_{\nu=1}^p \Gamma(L-\nu+1)$.

The Hellinger distance between two scaled complex Wishart distributions with the same number of looks $L_{i}=L_{j}=L$ is given by
\begin{equation}
d_{H}(\mathbf{\Sigma}_{i}, \mathbf{\Sigma}_{j})  = 1 - 
	\bigg[	
	\frac{
		\big| \big(
			\mathbf{\Sigma}_{i}^{-1} + \mathbf{\Sigma}_{j}^{-1} 
		\big)^{-1} \big|
		}{
			2 \sqrt{
				|\Sigma_{i}| |\Sigma_{j}|}
		}
	\bigg]^L,
\label{eq:hellinger_multidimen}
\end{equation}
where we indexed the distributions by their parameters.
Consequently the Hellinger distance for the 1-dimensional ($p=1$) multilook intensity distribution is given by
\begin{equation}
d_{H}(\sigma_{i}^{2}, \sigma_{j}^{2}) = 1 - {\bigg[
\frac{\sigma_{i} \sigma_{j}}{{2(\sigma_{i}^{2}}+{\sigma_{j}^{2})}}
\bigg]}^{L}.
\label{eq:hellinger_onedimemsion}
\end{equation}
Expression~\eqref{eq:hellinger_onedimemsion} is the Hellinger distance between two Gamma distributions with same shape parameter $L$ and expected values $\sigma_{i}^{2}$ and $\sigma_{j}^{2}$. 
In the context of PolSAR data we can associate $\sigma_{i}^{2}$ and $\sigma_{j}^{2}$ of the two probability density functions with the diagonal elements $T_{\ell\ell}$, $T_{\ell\ell}(\theta)$ for $\ell=1,2,3$ of the un-rotated and the rotated Hermitian positive definite coherency matrices respectively; cf.~\eqref{eq:orienting_T_matrix}. 
%%% ACF Actually, $\sigma_{1}^{2}$ and $\sigma_{2}^{2}$ are diagonal entries of the covariance matrix

\section{Polarization Orientation Angle Estimation}\label{Sec:OAEstimation}

In this section we estimate the polarization OA from full-polarimetric SAR data by using a stochastic distance measure between the elements of the coherency matrix, which is assumed to follow a complex Wishart distribution.

The OA is zero for reflection symmetric media, but OA shifts are induced for surfaces with azimuthal tilts and for buildings oriented perpendicular to the radar LOS. 
Apart from these, the OA is appreciable for low frequency (L and P band) data in forested areas due to surface topography. 
In this context, a recent adaptive-model based decomposition with topographic polarization orientation compensation (TPOC) has been proposed in~\cite{Arii2012_IGARSS}. 
In this, the volume orientation is removed from the generalized volume component, and the conventional adaptive model-based decomposition is modified by introducing the TPOC concept. 

In general, the primary effect of compensating the OA is in the reduction of the cross-polarization $(T_{33})$ component and the increase in the co-polarized $(T_{22})$ component. 
Following the proposal of Lee and Ainsworth~\cite{Lee2011}, the Hermitian positive definite coherency matrix $\mathbf{[T]}$ is unitarily rotated by $\mathbf{[U_{3}]}$:
\begin{equation}
\begin{split}
\mathbf{[T(\theta)]} &= \mathbf{[U_{3}]}\mathbf{[T]}{\mathbf{[U_{3}]^{-1}}}, \\
\mathbf{[T]} &= \left[ \begin{array}{ccc}
T_{11} & T_{12} & T_{13} \\
T_{12}^{*} & T_{22} & T_{23} \\
T_{13}^{*} & T_{23}^{*} & T_{33}
\end{array}\right] 
=\left[ \begin{array}{ccc}
\sigma_{1}^{2} & \rho_{12} & \rho_{13} \\
\rho_{12}^{*} & \sigma_{2}^{2} & \rho_{23} \\ 
\rho_{13}^{*} & \rho_{23}^{*} & \sigma_{3}^{2}
\end{array}\right], \\
\mathbf{[U_{3}]} &= \left[ \begin{array}{ccc}
1 & 0 & 0 \\
0 & \cos(2\theta) & \sin(2\theta) \\
0 & -\sin(2\theta) & \cos(2\theta)
\end{array}\right].
\end{split}
\label{eq:orienting_T_matrix}
\end{equation}
The orientation angle $\theta\in[-{\pi}/{8},{\pi}/{8}]$ is estimated by minimizing the $T_{33}(\theta)$ component, \emph{i.e.}, ${dT_{33}(\theta)}/{d\theta}=0$, with this:
\begin{equation}
2\theta = \frac{1}{2}\tan^{-1} \frac{-2\Re (T_{23})}{T_{33}-T_{22}}.
\label{eq:lee_angle}
\end{equation}
The effect of orientation on the three diagonal elements of the coherency matrix is as follows: 
\begin{enumerate}
\item $T_{11}={\left|\mathbf{HH + VV}\right|^{2}}/{2}$ is roll invariant for any $\theta$, 
\item $T_{22}={\left|\mathbf{HH - VV}\right|^{2}}/{2}$ always increases or remains the same after the OA compensation, and
\item $T_{33}=2\left|\mathbf{HV}\right|^{2}$ always decreases or remains the same after OA compensation. 
\end{enumerate}

In this work, we have used a statistical information theoretic measure to estimate the OA from full-polarimetric SAR imagery: 
the OA is estimated by maximizing the Hellinger distance between the $T_{33}$ and the $T_{22}$ elements of $\mathbf{[T]}$. 

The proposed method estimates the OA by first maximizing the Hellinger distance between the un-rotated ($\sigma_{2}^{2}$, $\sigma_{3}^{2}$) and the rotated ($\sigma_{2}^{2}(\theta)$, $\sigma_{3}^{2}(\theta)$) elements, over the $[-{\pi}/{4},{\pi}/{4}]$ range, leading to two candidate angles:
\begin{equation}
\begin{aligned}
\phi_{3} & = \argmax_{-\pi/4\le\theta\le \pi/4}\left\{ 1-\left[ \frac{2\sqrt{\sigma_{3}^{2}\sigma_{3}^{2}(\theta)}}{\sigma_{3}^{2}+\sigma_{3}^{2}(\theta)}\right]^{L} \right\}, \text{ and}\\
\phi_{2} & = \argmax_{-\pi/4\le\theta\le \pi/4}\left\{ 1-\left[ \frac{2\sqrt{\sigma_{2}^{2}\sigma_{2}^{2}(\theta)}}{\sigma_{2}^{2}+\sigma_{2}^{2}(\theta)}\right]^{L} \right\}.
\end{aligned}
\end{equation}
Two maxima are found at $\phi = \phi_{\{3,2\}}$ and $\phi = \phi_{\{3,2\}}\pm \pi/4$, and the OA is chosen such that the Hellinger distance ($d_{H\phi_{3}}$) corresponding to $\sigma_{3}^{2}$ is greater than the Hellinger distance ($d_{H\phi_{2}}$) corresponding to $\sigma_{2}^{2}$ either at $\phi = \phi_{\{3,2\}}$ or at $\phi = \phi_{\{3,2\}}\pm \pi/4$. 
This condition corresponds exactly to the case mentioned earlier, where the cross-polarized component is minimized, whereas the other peak corresponds to the situation where the cross-polarized component is incorrectly maximized. 
Finally, the OA $(\theta_{0})$ is obtained by wrapping $\phi$ in the range $[-{\pi}/{8},{\pi}/{8}]$ using~\eqref{eq:final_OA}, which is then compared with the OA obtained by the method stated in~\cite{Lee2011}:
\begin{equation}
\theta_{0}=
\begin{cases}
\phi+\pi/4 & \text{if}\  \phi<-\pi/8, \\
\phi-\pi/4 & \text{if}\  \phi>\pi/8, \\
\phi & \text{otherwise.}
\end{cases}
\label{eq:final_OA}
\end{equation}

The implementation of the proposed method is given in Algorithm~\ref{alg:the_alg}.
The code is freely available at \url{https://github.com/avikcsre/SD_Y4O/}.

\begin{figure*}[hbt]
\centering
\subfloat[]{\includegraphics[width=0.591\columnwidth]{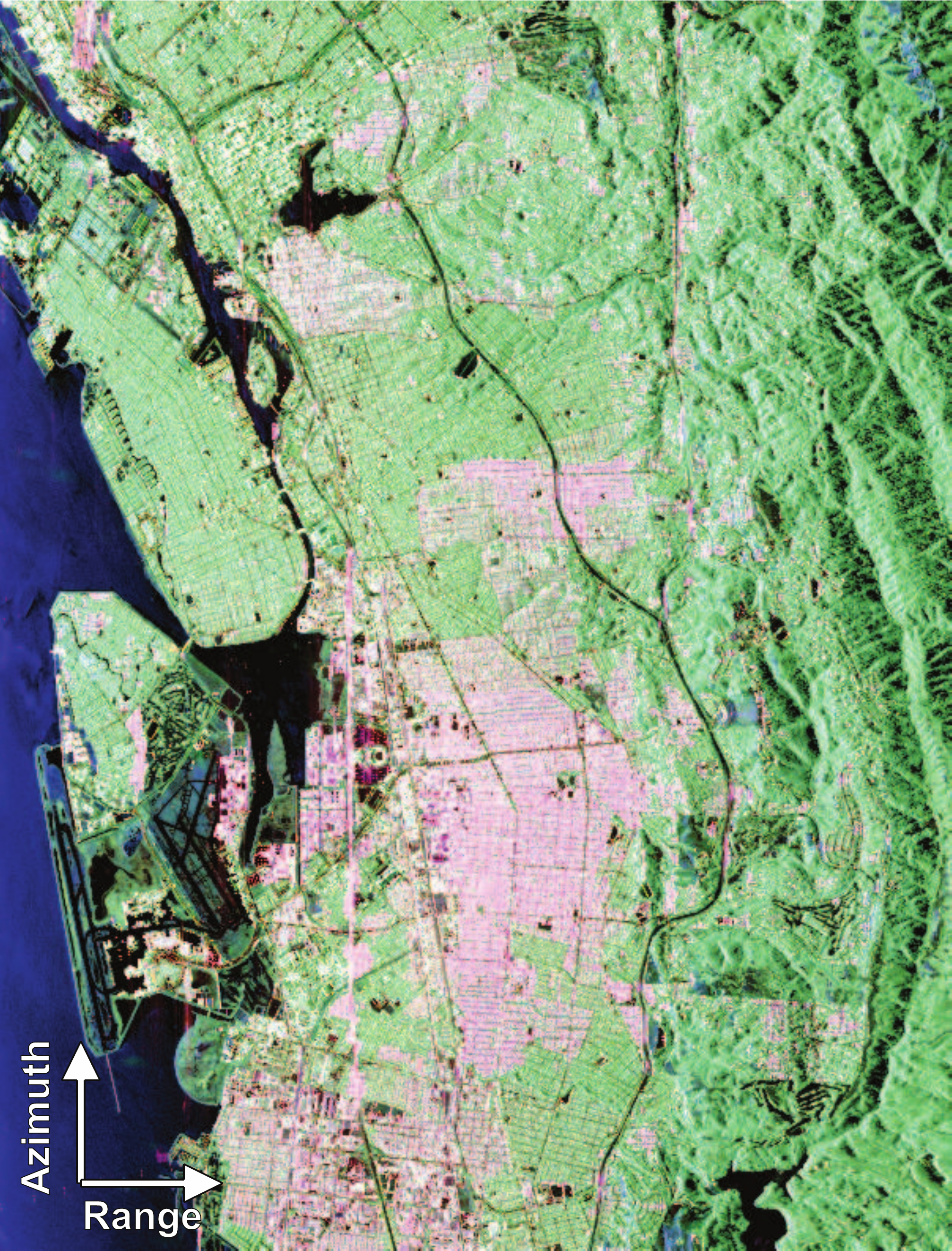}} \hspace{1mm}
\subfloat[]{\includegraphics[width=0.596\columnwidth]{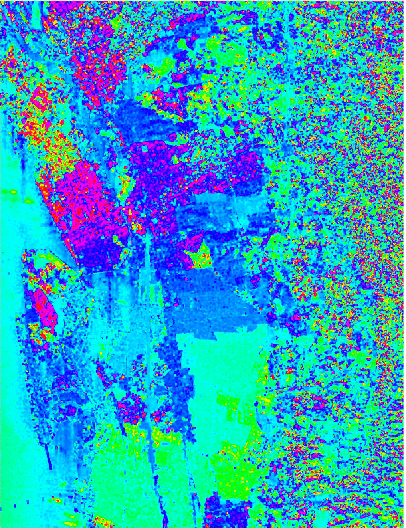}} \hspace{1mm}
\subfloat[]{\includegraphics[width=0.7\columnwidth]{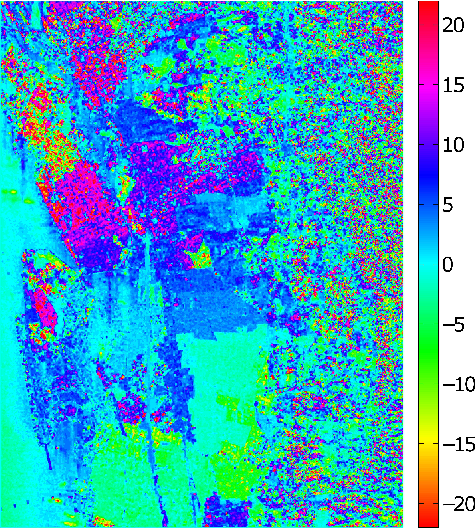}}
\caption{(a). UAVSAR L-band Pauli RGB image of the Hayward area, (b). OA estimated by the method proposed in~\cite{Lee2011} and (c). OA estimated by the proposed method using the Hellinger distance.}
\label{fig:theory_results}
\end{figure*}

\begin{algorithm}[hbt]
\label{alg:the_alg}
\caption{Orientation Angle Estimation}
\KwData{\boldmath{$[T]$} matrix}
\KwResult{Orientation compensating angle ($\theta_{0}$)}
\Begin{
 $\mathbf{[T]}^\prime \leftarrow \mathbf{[U_{3}(\theta)]} \mathbf{[T]}\mathbf{[U_{3}(\theta)]}^{-1} \  
 \forall\  \theta \rightarrow [-\pi/4 , +\pi/4] $ \;
 Find common peaks in $d_{H\phi_3} \leftarrow d_{H}(\sigma_3^2,\sigma_3^2(\theta))$ and $d_{H\phi_2} \leftarrow d_{H}(\sigma_2^2,\sigma_2^2(\theta))$ \;
 \BlankLine
 At $\phi_{\{3,2\}}$:
  \eIf{ $d_{H\phi_{3}} > d_{H\phi_{2}}$}{
   Select $\phi \leftarrow \phi_3$ \;}{Select $\phi \leftarrow \phi_2$\;}
   \BlankLine
   \uIf{$\phi > {\pi}/{8}$}{$\theta_{0} \leftarrow \phi - {\pi}/{4} $ \;}
   \uElseIf{$\phi < -{\pi}/{8}$}{$\theta_{0} \leftarrow \phi + {\pi}/{4}$ \;}
   \Else{$\theta_{0} \leftarrow \phi$ \;}
   }
\end{algorithm}

\begin{figure}[hbt]
\centering
\includegraphics[width=1\columnwidth]{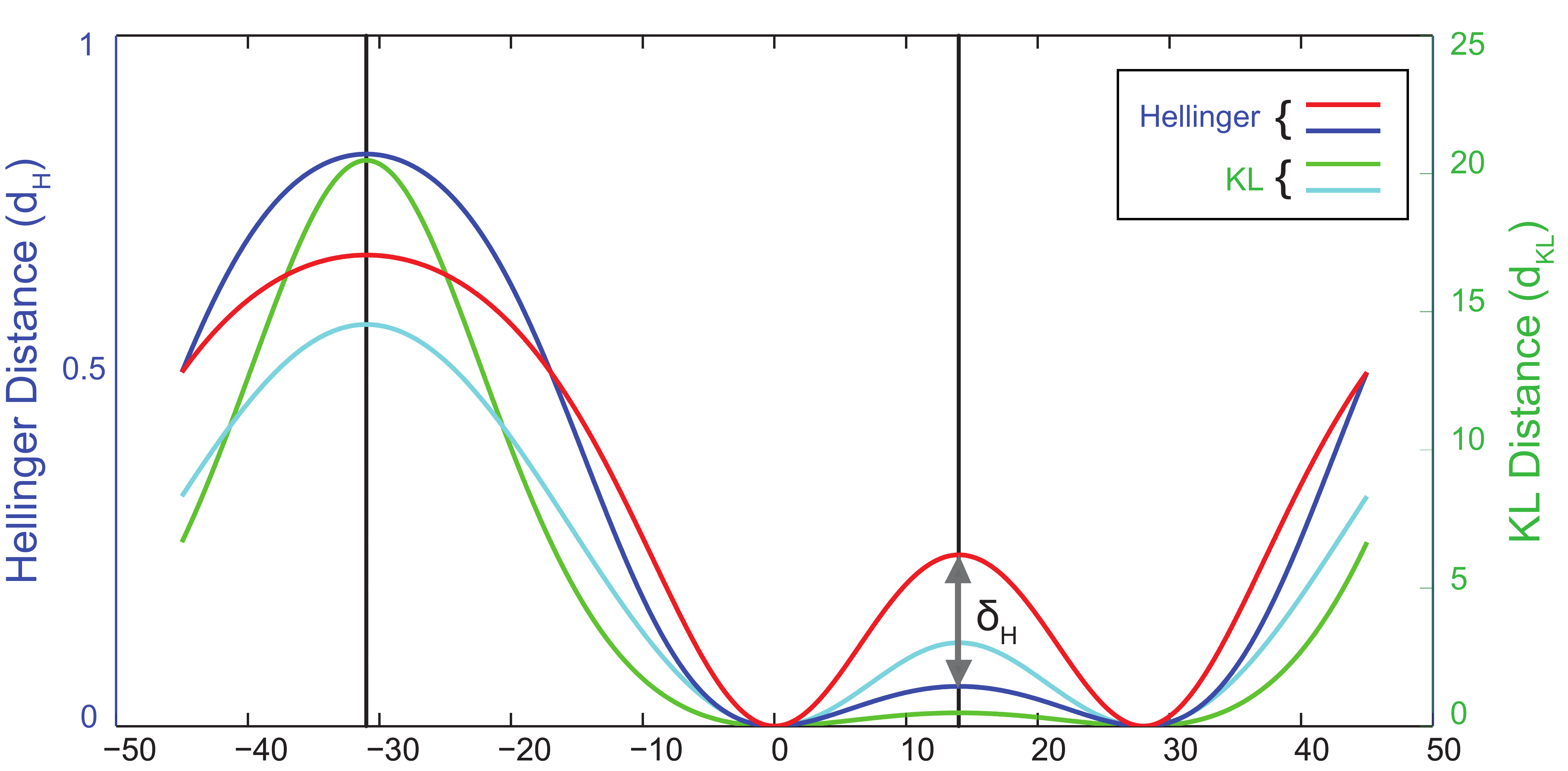}
\caption{Variation of Hellinger distances ($d_{H}(\sigma_{3}^{2}, \sigma_{3}^{2}(\phi))$ in red and $d_{H}(\sigma_{2}^{2}, \sigma_{2}^{2}(\phi))$ in blue) and KL distances ($d_{KL}(\sigma_{3}^{2}, \sigma_{3}^{2}(\phi))$ in cyan and $d_{KL}(\sigma_{2}^{2}, \sigma_{2}^{2}(\phi))$ in green) with $\phi$; $\delta_{H}$ is the relative Hellinger distance between $T_{33}$ and $T_{22}$ at $\phi=\theta_{0}$.}
\label{fig:hellinger_ex_plot}
\end{figure}

In the following we use a coherency matrix $\mathbf{[T]}$ extracted from an urban area, to illustrate the proposed methodology:
\begin{equation}
\mathbf{[T]}=\left[ \begin{array}{ccc}
4.56 & 2.28+0.72i & 0.02+0.67i \\
2.28-0.72i & 6.06 & 1.90+0.27i \\
0.02-0.67i & 1.90-0.27i & 3.50
\end{array}\right].
\label{eq:ex_T_matrix}
\end{equation}
The maxima estimated by the proposed methodology are, approximately, at $\phi=\ang{-31}$ and at $\phi=\ang{14}$, as shown in Fig.~\ref{fig:hellinger_ex_plot}. 
The orientation compensating angle at $\theta_{0}=\ang{14}$ is chosen because it satisfies the aforementioned criterion of $d_{H}(\sigma_{3}^{2}, \sigma_{3}^{2}(\phi)) \ge d_{H}(\sigma_{2}^{2}, \sigma_{2}^{2}(\phi))$. 

Apart from the Hellinger distance, any other distances mentioned in Section~\ref{Sec:StochasticDistances} can be used to estimate the OA. 
An example is shown with the Kullback-Leibler (KL) distance in Fig.~\ref{fig:hellinger_ex_plot} to estimate the OA. 
It can be seen that the OA is also correctly estimated by using the KL distance for the coherency matrix given in~\eqref{eq:ex_T_matrix}. 
Our choice was based on computational cost, since the Hellinger distance is the least expensive among the available ones. 

\section{SD-Y4O}\label{Sec:sd-y4o}

% %AB: The paragraph that you had marked is changed and modified here
Presently, all the decomposition techniques that account for the overestimation of the volume power due to target orientation~\cite{Lee2011,An10,YAMAGUCHI2011,singh13,Arii11,Chen14} do so either by rotating the coherency matrix or by considering different volume scattering models. 

In this work, a modification of the scattered powers for the Yamaguchi four component decomposition (Y4O) is obtained using the relative Hellinger distance. 
Fig.~\ref{fig:hellinger_ex_plot} shows the relative distance $\delta_{H}$ between the two Hellinger distances $d_{H}(\sigma_{3}^{2}, \sigma_{3}^{2}(\phi))$ and $d_{H}(\sigma_{2}^{2}, \sigma_{2}^{2}(\phi))$ computed at $\phi$ (the unwrapped orientation angle in the $[-{\pi}/{4},{\pi}/{4}]$ range).
This relative distance is used to modify the Y4O scattering powers. 

According to Lee and Ainsworth~\cite{Lee2011}, the amount of increase in double-bounce power is not equal to the amount of decrease in the volume power. 
It has been shown that the amount of decrease in the volume power component is usually greater than the amount of increase in the double-bounce power component. 
Moreover, the increase in surface power is smaller compared to the increase in double-bounce power. 
Due to the roll-invariant property, the helix power remains unaltered after rotation. 

In the following we quantify the effect of the proposed rotation on the decomposition powers of a particular target:  
\begin{equation}
0 \le \delta_{H}=d_{H}(\sigma_{3}^{2}, \sigma_{3}^{2}(\phi)) - d_{H}(\sigma_{2}^{2}, \sigma_{2}^{2}(\phi)) \le 1.
\label{eq:OA_threshold}
\end{equation}

\begin{figure}[hbt]
\centering
\includegraphics[width=1\columnwidth]{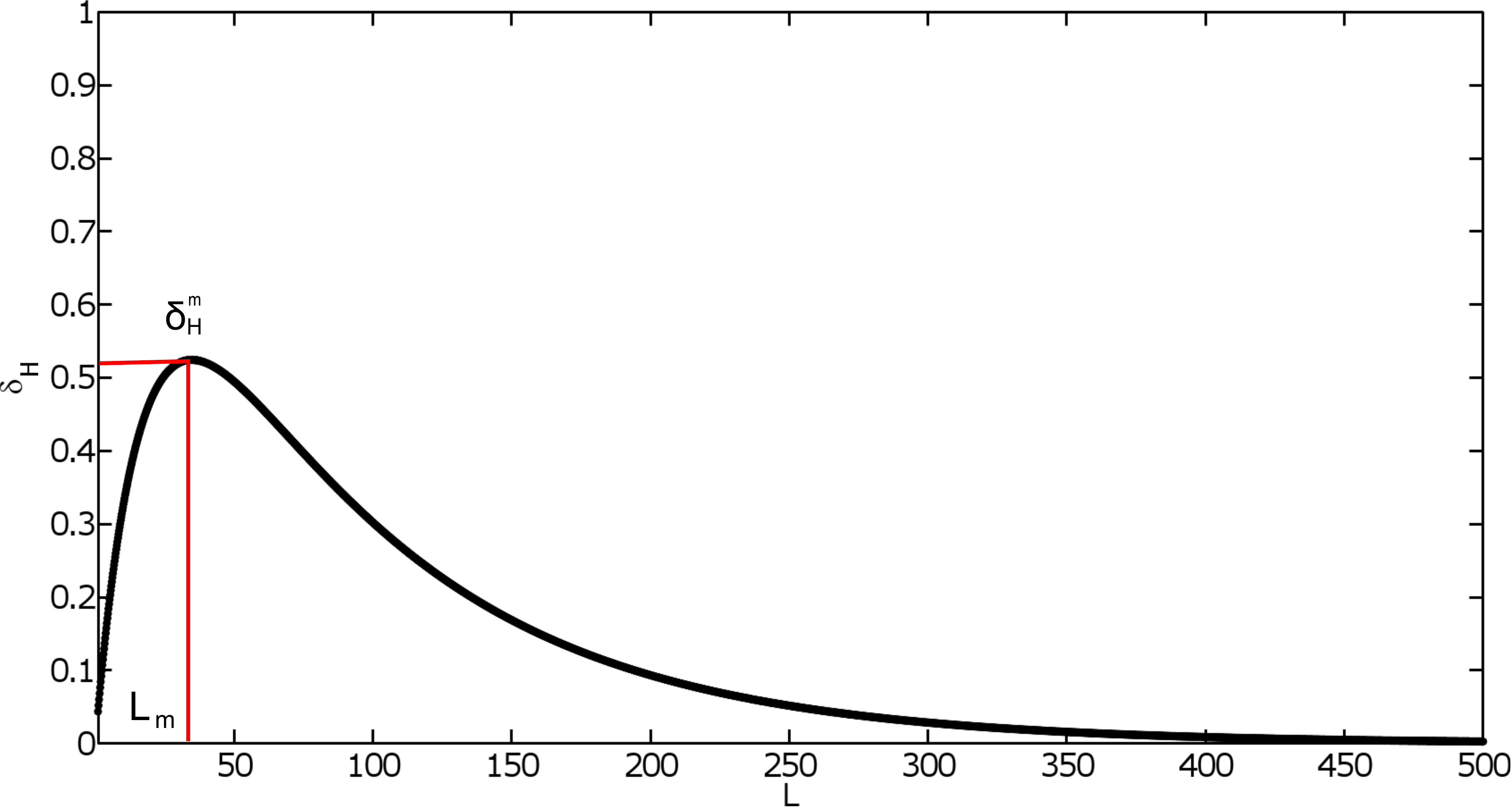}
\caption{The variation of $\delta_{H}$ with $L$ for $\phi=14^\circ$. A maximum of $\delta_H=\delta_H^m$ is obtained for a certain value of $L=L_m$.}
\label{fig:deltaH_L}
\end{figure}

From the previous section, it can be observed that the OA estimation is independent of $L$, but the relative distance $\delta_{H}$ increases with increasing $L$ until a maximum is attained, and then it rapidly decreases to zero as shown in Fig.~\ref{fig:deltaH_L}.
%%% ACF What follows was not clear; please check if it expresses what was done.
The Y4O decomposition powers are then modified with the estimated $\delta_{H}=\delta_{H}^{m}$ for $L=L_{m}$, where $\delta_{H}^{m}$ is the maximum attainable divergence between $\sigma_{3}$ and $\sigma_{2}$.

In the proposed methodology (SD-Y4O), we modify the Yamaguchi four component decomposition powers based on the maximum relative Hellinger distance $(\delta_{H}^{m})$ by:
\begin{subequations}
\begin{align}
P_{v}^{n} &= P_{v}(1-\delta_{H}^{m}), \\
P_{d}^{n} &= P_{d} + \alpha P_{v}\delta_{H}^{m} ,\\ 
P_{s}^{n} &= P_{s} + \beta P_{v}\delta_{H}^{m}, \text{ and} \\
P_{c}^{n} &= P_{c},
\end{align}
\end{subequations}
where $P_{v}$, $P_{d}$, $P_{s}$ and $P_{c}$ are the volume, double-bounce, surface and helix power components from the Yamaguchi four-component decomposition (Y4O), respectively. 
The new four-component powers: $P_{v}^{n}$, $P_{d}^{n}$ and $P_{s}^{n}$ are obtained by either deducting or adding the volume power component $(P_{v})$ of the Y4O decomposition, adjusted by the relative distance $\delta_{H}^{m}$ respectively. 
%%% ACF I am not sure if the rest of the paragraph belongs to this point of the manuscript

\begin{figure}[hbt]
\centering
\subfloat[$\alpha$]{\includegraphics[width=0.5\columnwidth]{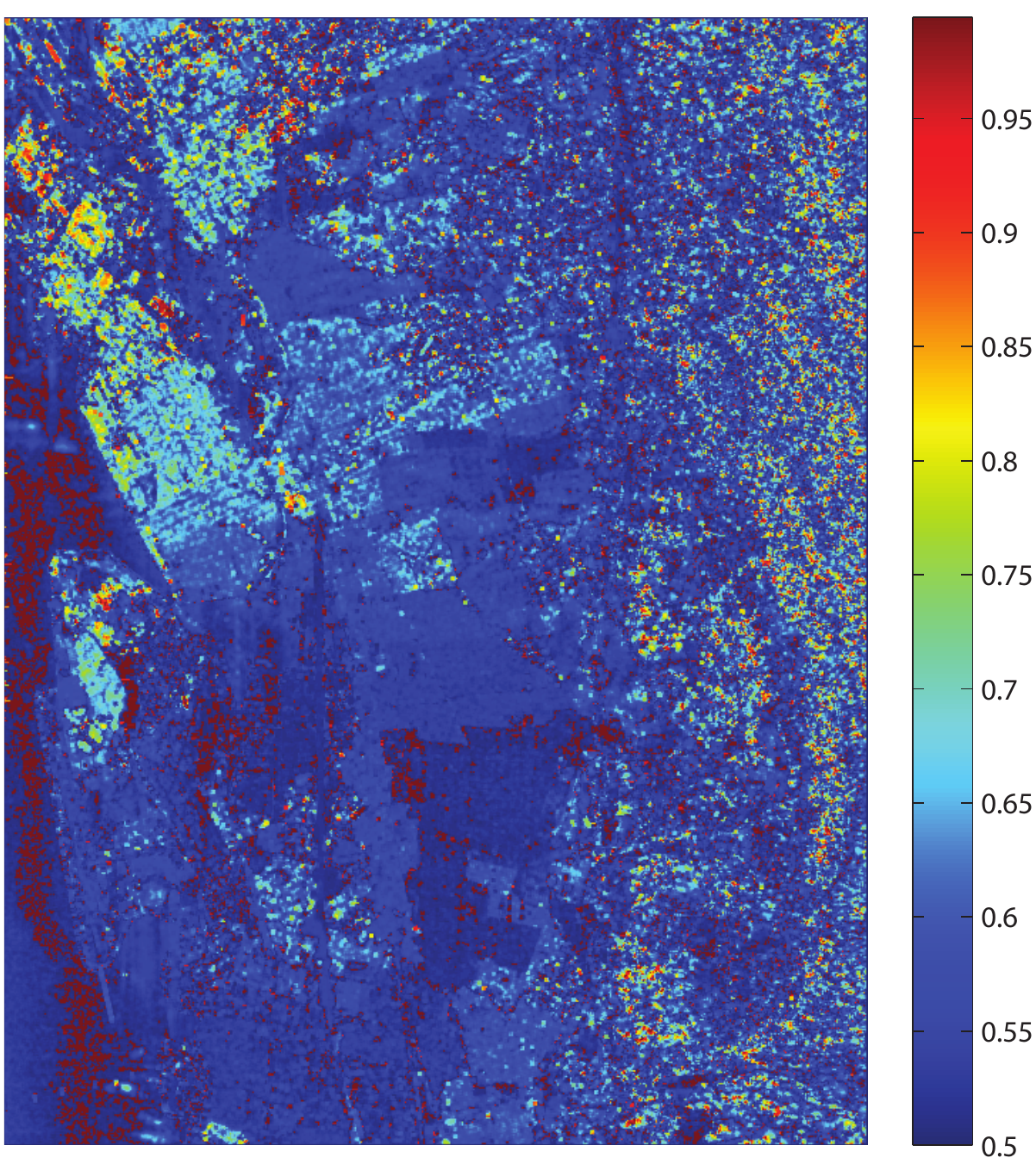}} 
\subfloat[$\beta$]{\includegraphics[width=0.5\columnwidth]{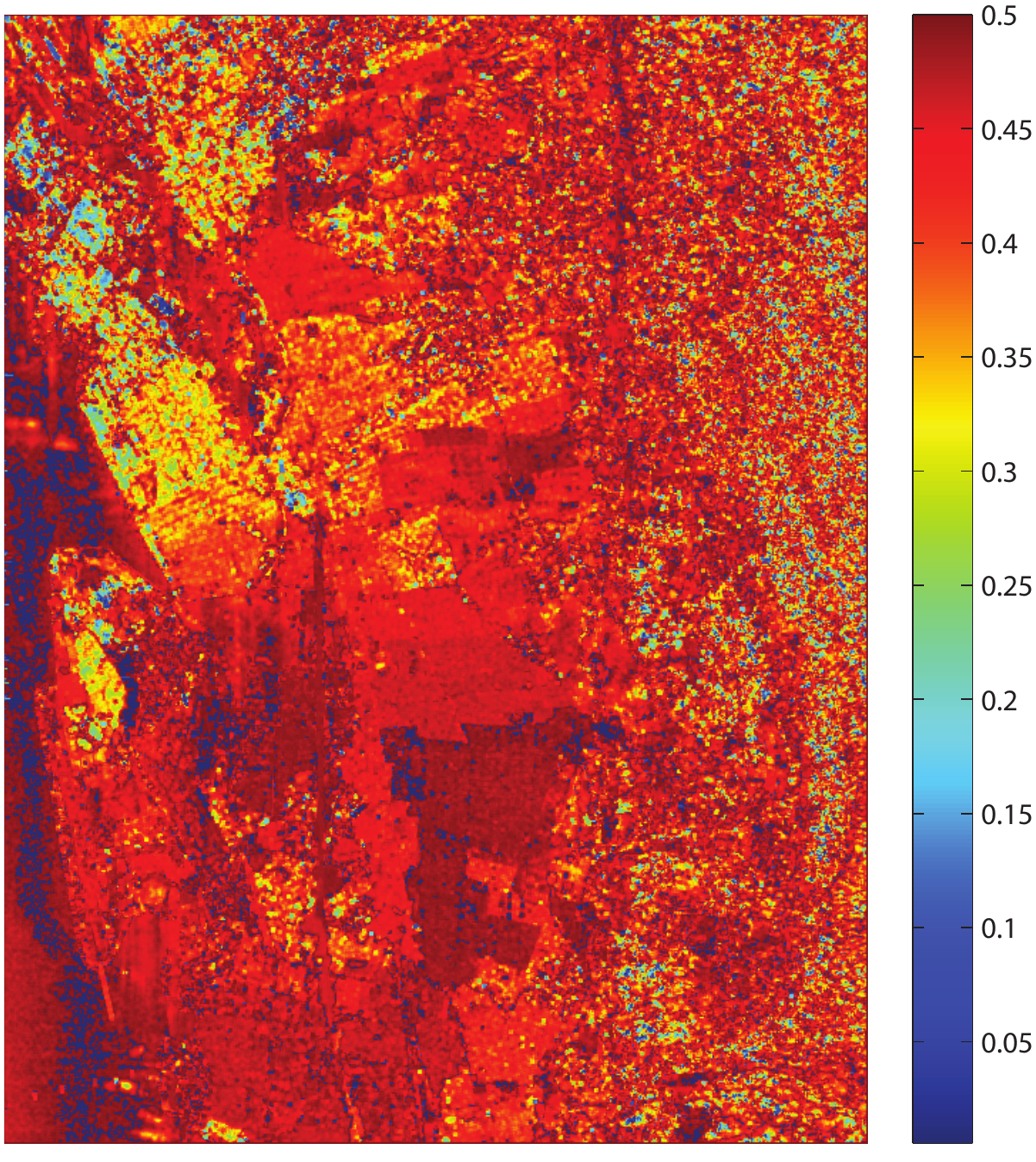}}
\caption{The $\alpha$ and the $\beta$ parameters maps for the UAVSAR L-band Hayward image.}
\label{fig:alpha_beta}
\end{figure}

The modulating factor $\delta_{H}^{m}$ is adjusted by the positive parameters $\alpha$ and $\beta$ with $\alpha+\beta=1$ for $P_{d}^{n}$ and $P_{s}^{n}$ respectively in order to conserve the total power ($TP=P_s+P_d+P_v$). 
Moreover, adhering to the criteria in~\cite{Lee2011}, \emph{i.e.,} the increase in the double-bounce scattering power is more than the increase in the surface scattering power, the ranges of the two parameters are set to $0.5\le\alpha\le 1$ and $0\le\beta\le 0.5$. 
The $\alpha$ parameter for each pixel is obtained by linearly mapping the estimated orientation angle ($\phi\in [0, {\pi}/{4}]$) to $[0.5, 1.0]$. 
The maps of the $\alpha$ and $\beta$ parameters for the UAVSAR L-band image are shown in Fig.~\ref{fig:alpha_beta}. 

\begin{figure}[hbt]
\centering
\subfloat[\label{Fig:SubFigImage}]{\includegraphics[width=0.6\columnwidth]{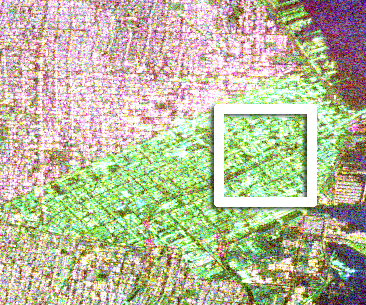}} \\
\subfloat[\label{Fig:SubFigAlpha}]{\includegraphics[width=0.5\columnwidth]{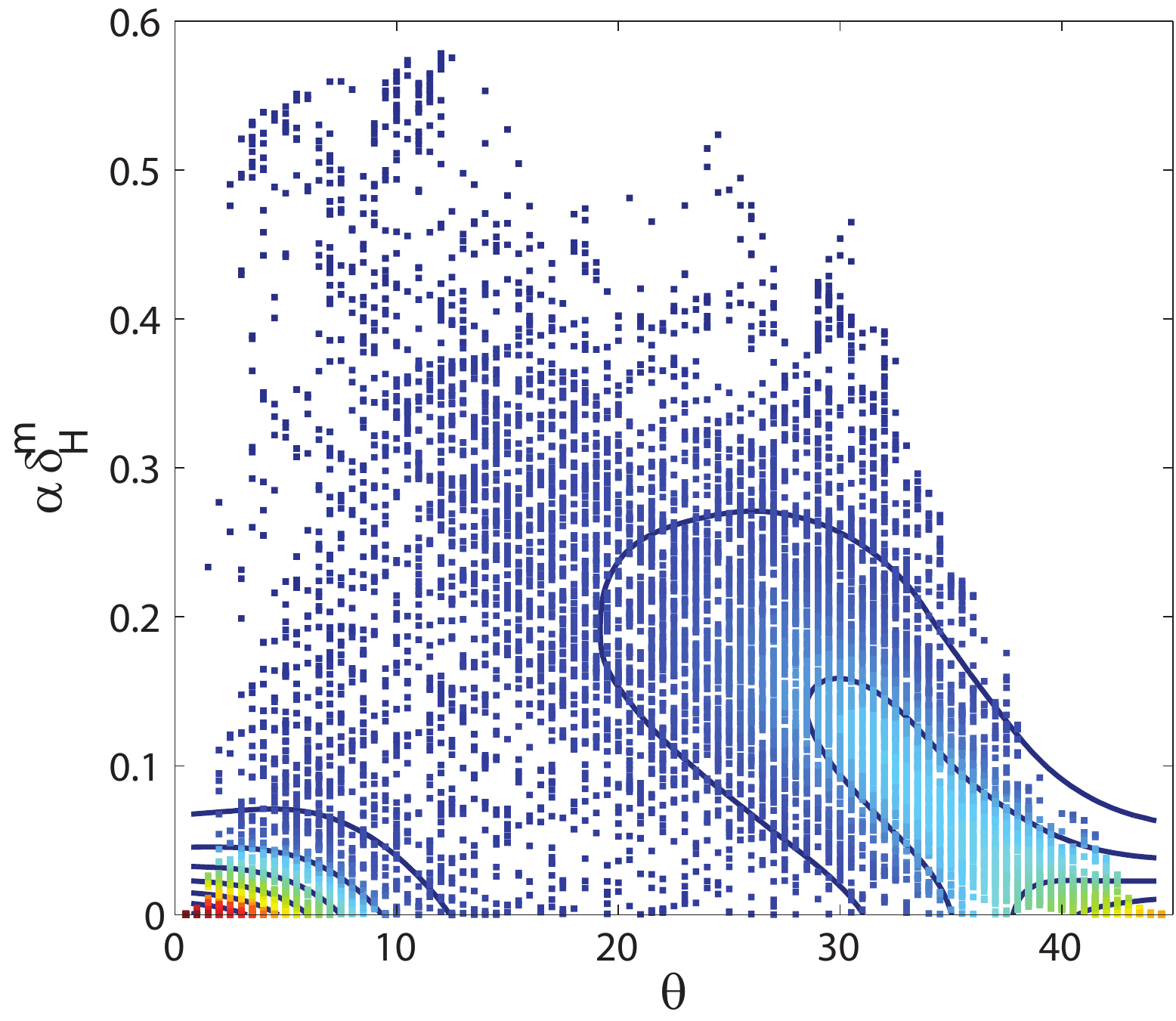}} 
\subfloat[\label{Fig:SubFigBeta}]{\includegraphics[width=0.5\columnwidth]{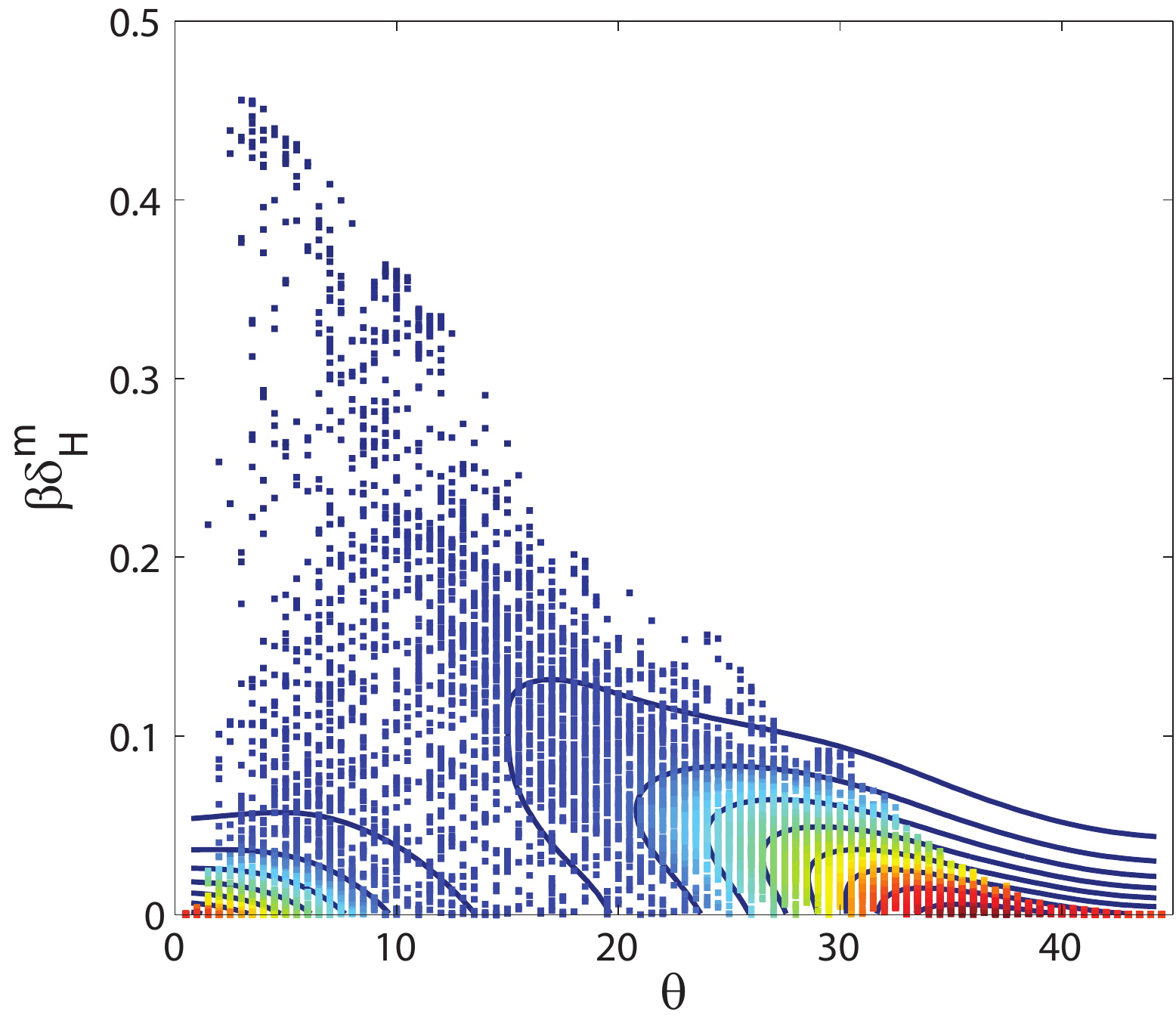}}
\caption{(a) The Pauli RGB of a oriented urban area, (b) and (c) the scatterer plot along with density showing the variation of $\alpha\delta_{H}^m$ and $\beta\delta_{H}^m$ with $\theta$}
\label{fig:results_1}
\end{figure}

It should be noticed that the linear mapping of the estimated OA to the  $\alpha$ parameter does not guarantee maximum $P_{d}$ (and $P_{s}$) powers to be observed at/near $45^\circ$. 
The $P_{d}$ and $P_{s}$ powers are modified by using the relative Hellinger distance $\delta_{H}^m$, which is a non-linear function of the OA angle, multiplied by the $\alpha$ parameter. 
Figs.~\ref{Fig:SubFigAlpha} and~\ref{Fig:SubFigBeta} show the 2d scatter plots of $\alpha\delta_{H}^m$, and $\beta\delta_{H}^m$ versus $\theta$, respectively, with the contours representing the density for oriented urban areas as shown in Fig.~\ref{Fig:SubFigImage}. 
It can be seen that the maximum density of points lies between $\theta=\ang{20}$ to $\theta=\ang{35}$. 

\begin{figure}[hbt]
\centering
\subfloat[]{\includegraphics[width=\columnwidth]{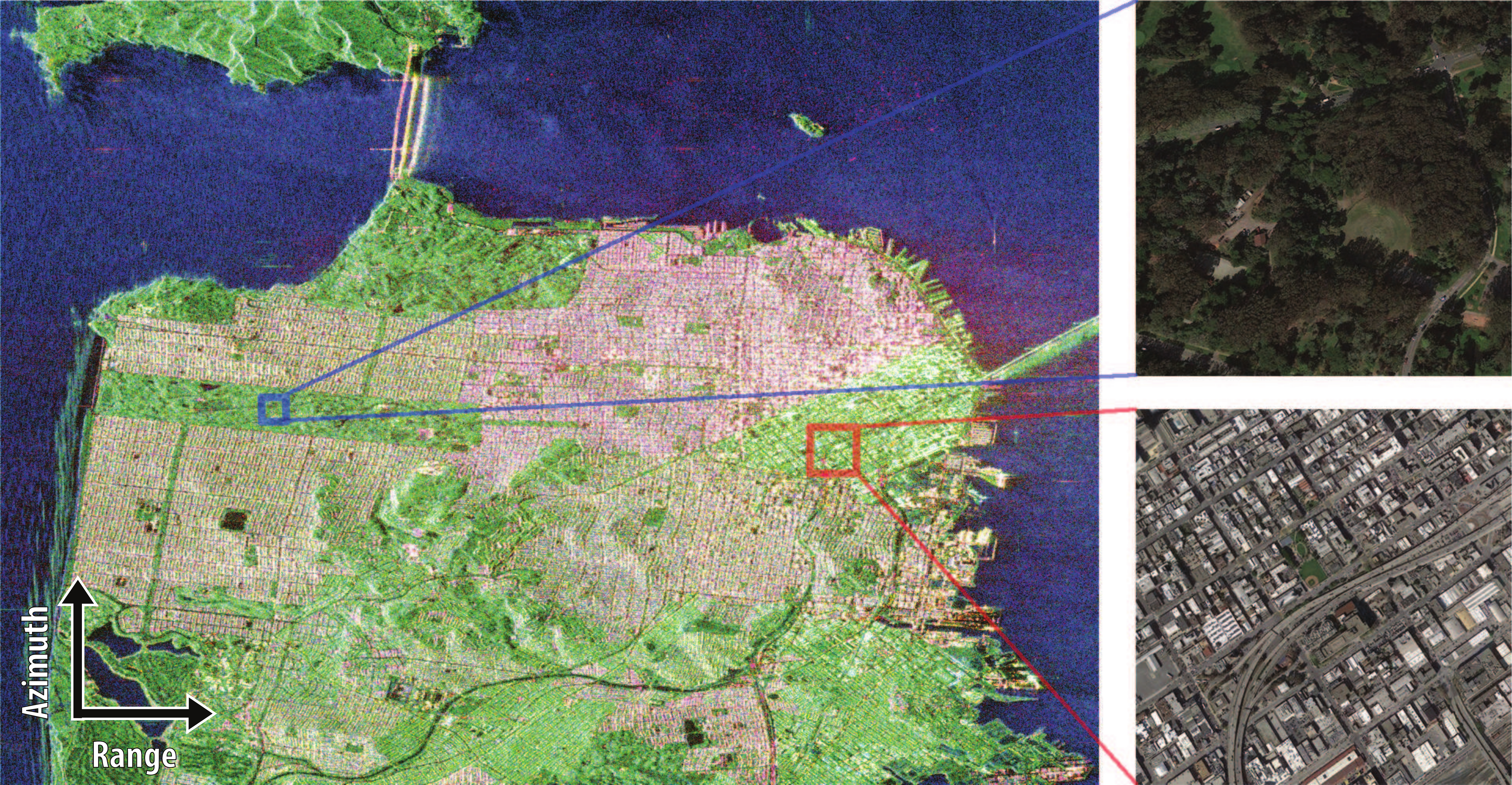}} \\
\subfloat[]{\includegraphics[width=\columnwidth]{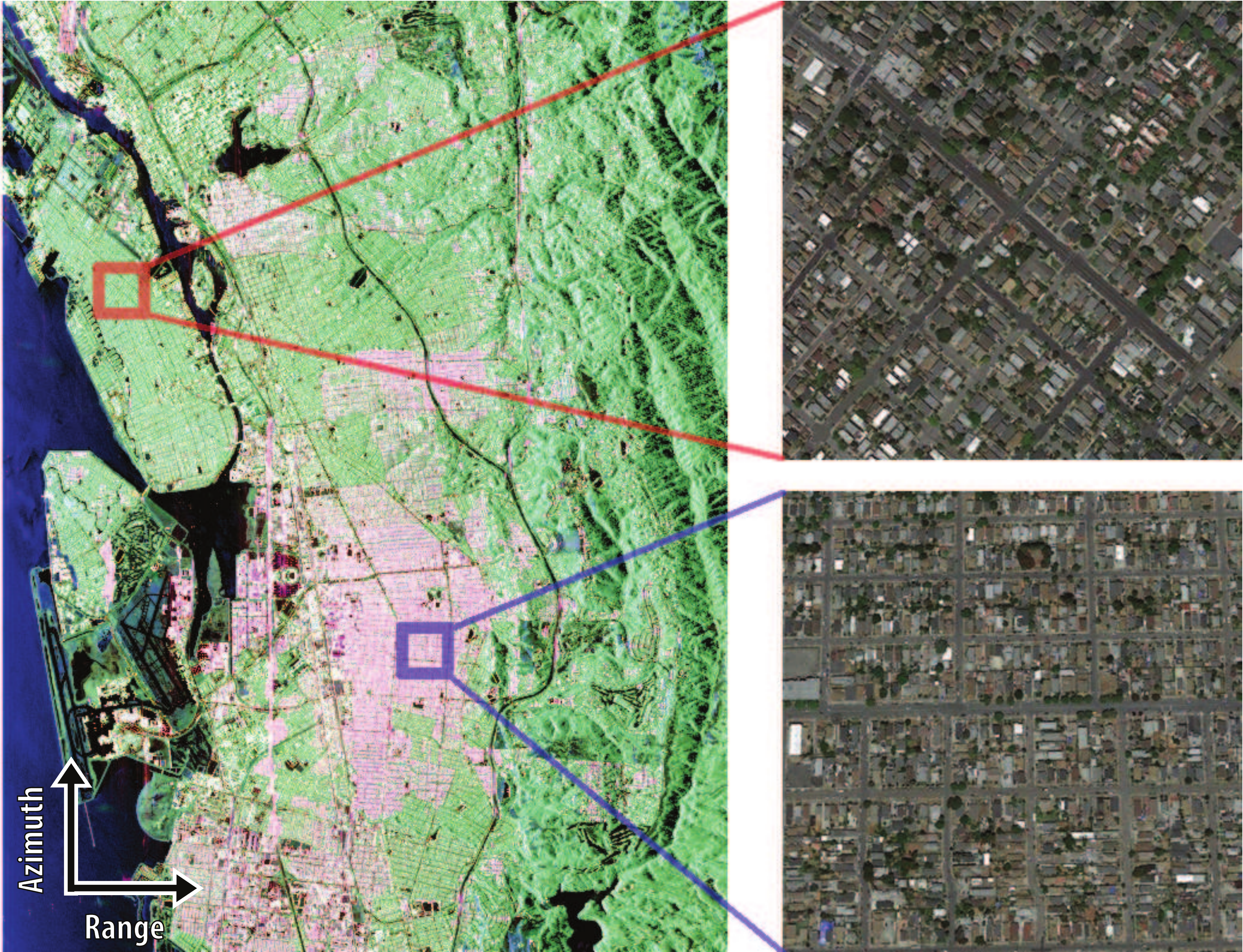}}
\caption{Pauli RGB of (a) Radarsat-2 data over San Francisco and (b) UAVSAR data over Hayward. Zoomed optical images are shown alongside for some areas. Courtesy: Google earth.}
\label{fig:SF_UAVSAR}
\end{figure}  

In the following section the SD-Y4O method is quantitatively compared with the Y4O and Y4R decomposition scattering powers for a Radarsat-2 C-band and a UAVSAR L-band.

\begin{figure*}[!h]
\centering
\subfloat[Y4O]{\includegraphics[width=0.65\columnwidth]{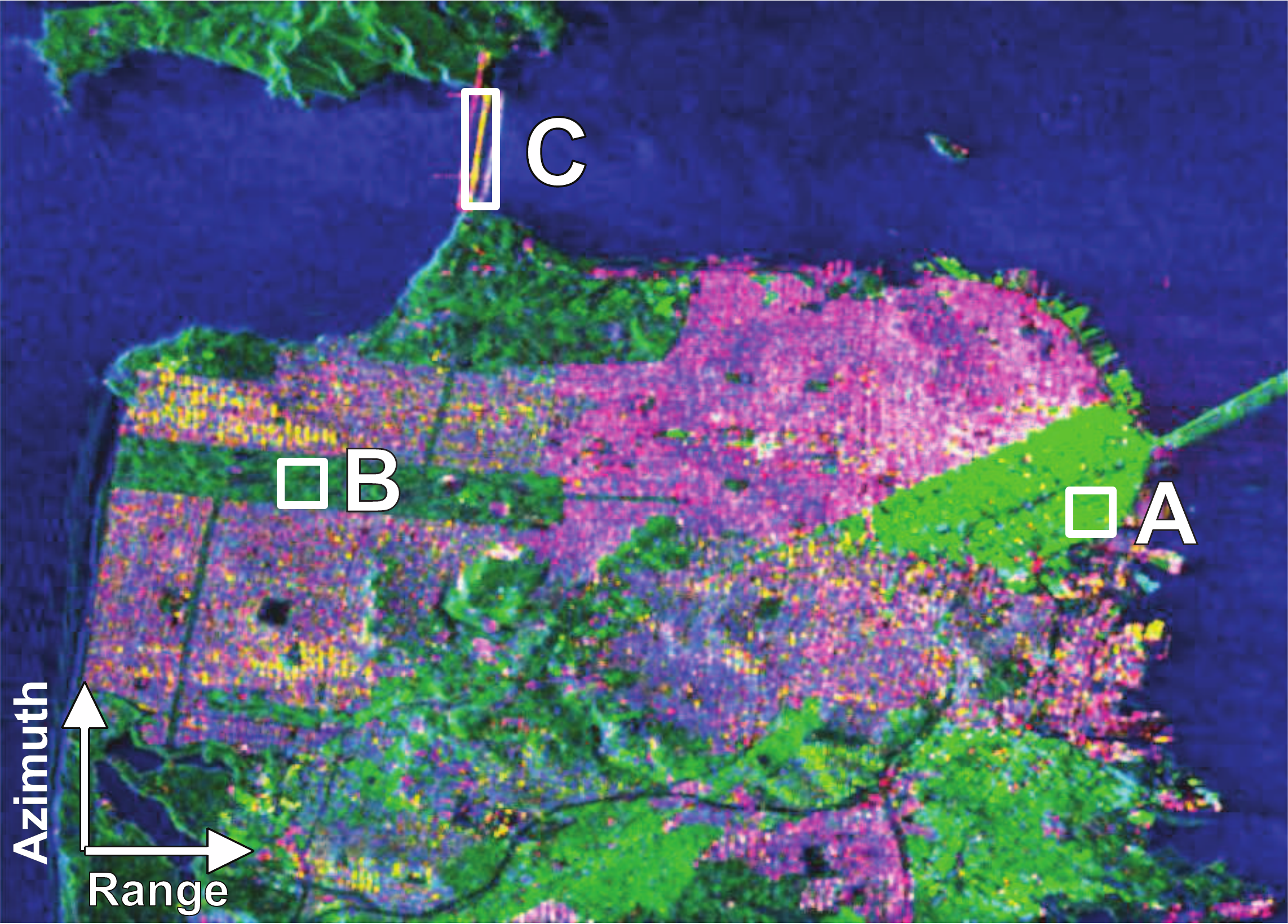}} \hspace{1mm}
\subfloat[Y4R]{\includegraphics[width=0.65\columnwidth]{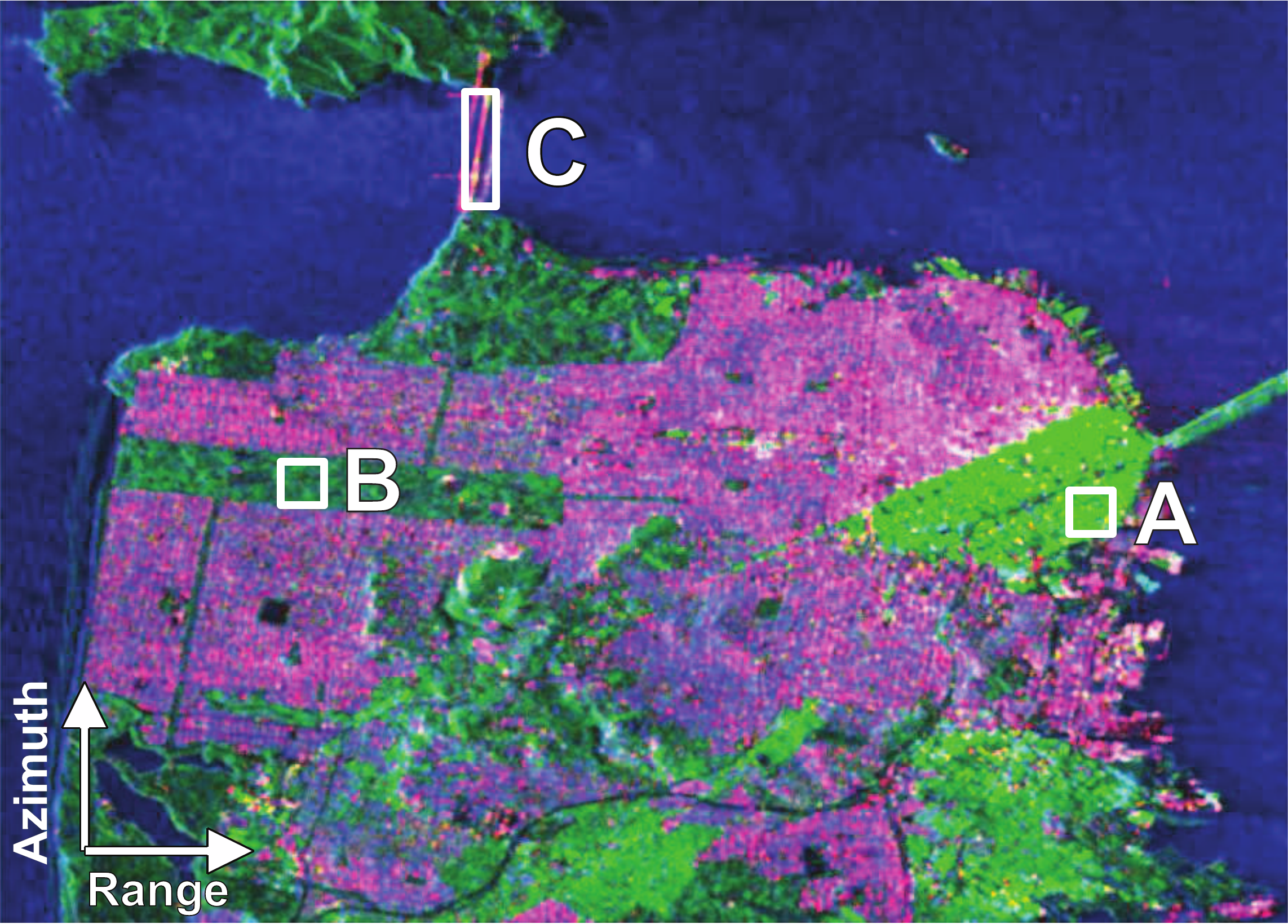}} \hspace{1mm}
\subfloat[SD-Y4O]{\includegraphics[width=0.65\columnwidth]{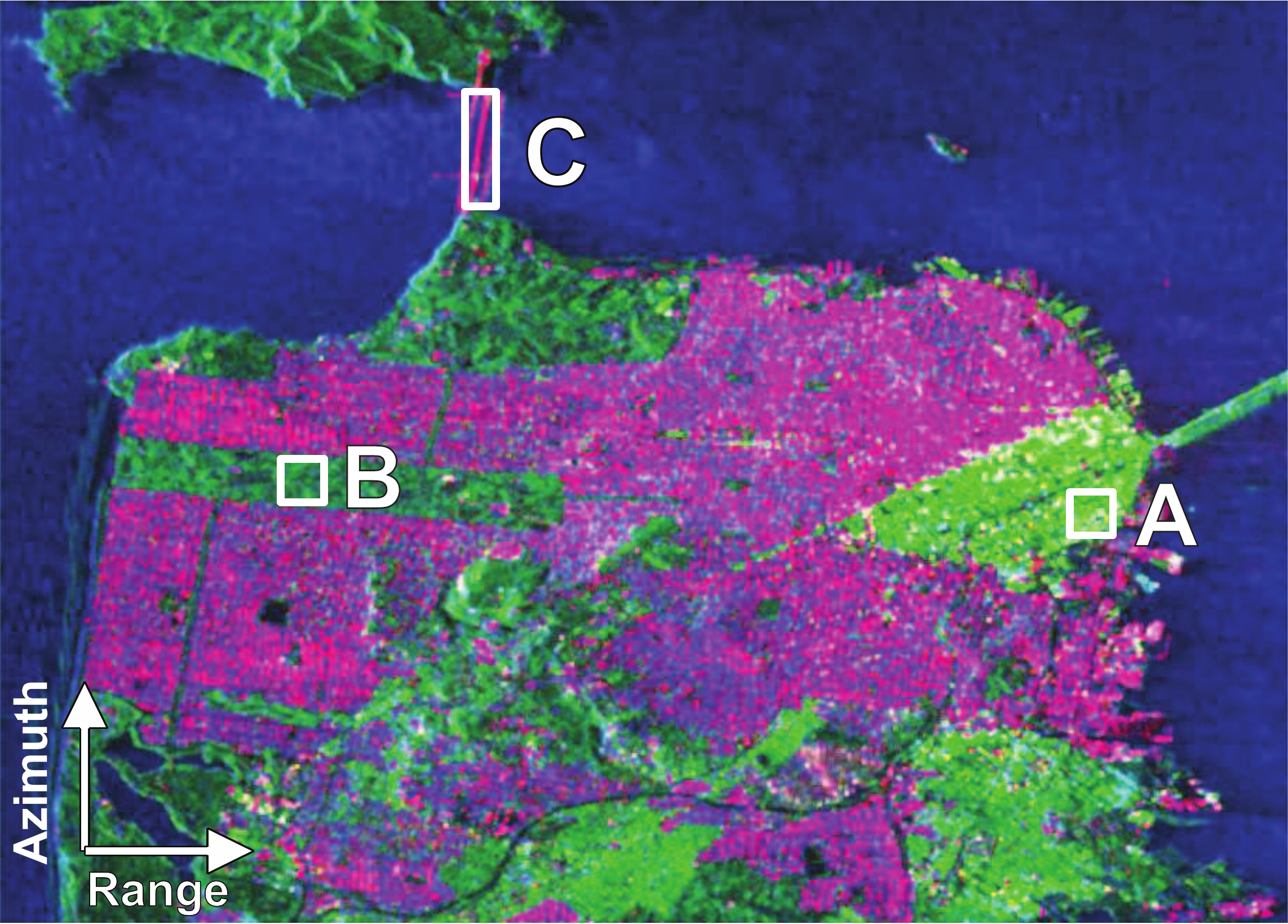}} \\
\subfloat[Y4O Area A]{\includegraphics[trim=8mm 8mm 8mm 8mm,clip,width=0.45\columnwidth]{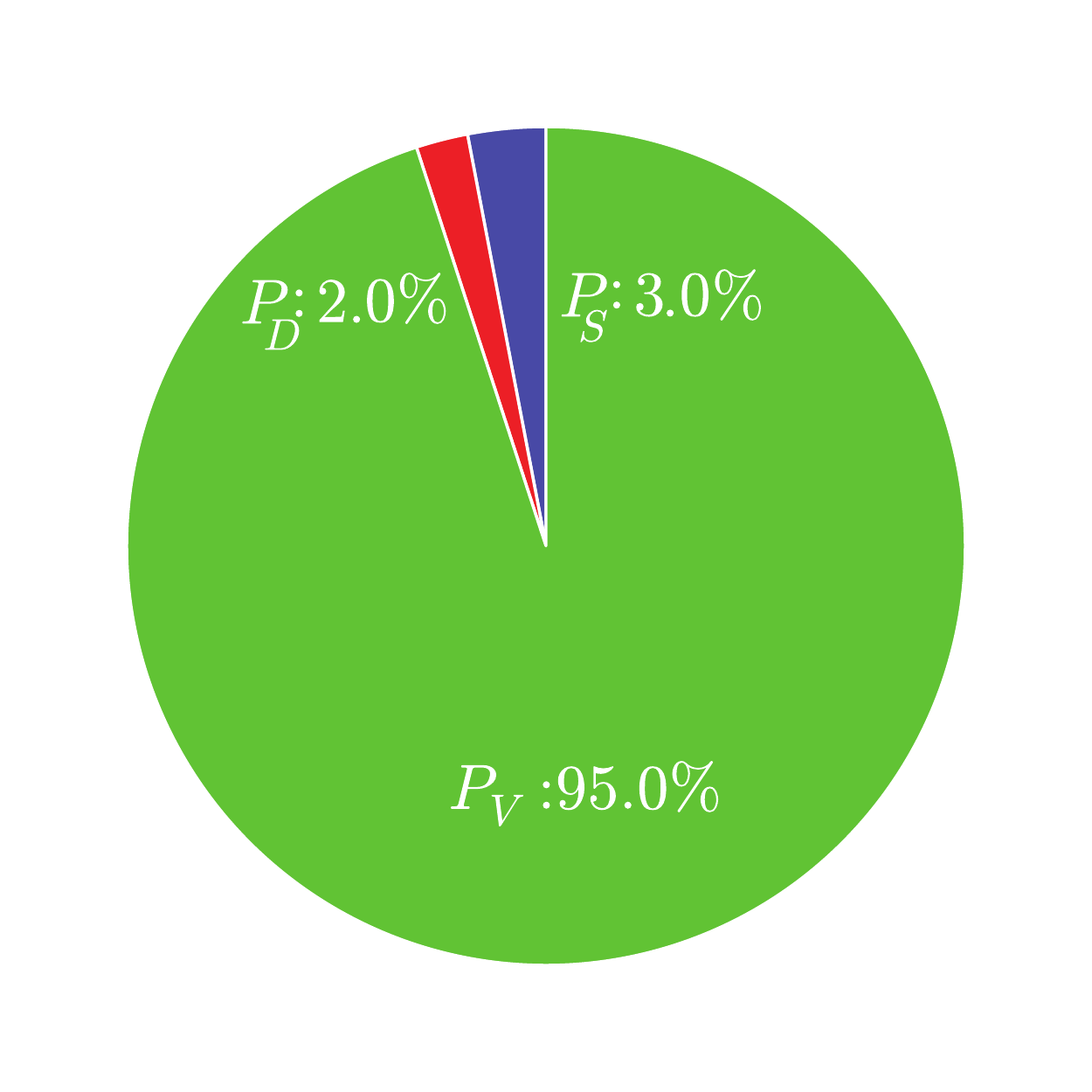}} \hspace{20mm}
\subfloat[Y4R Area A]{\includegraphics[trim=8mm 8mm 8mm 8mm,clip,width=0.45\columnwidth]{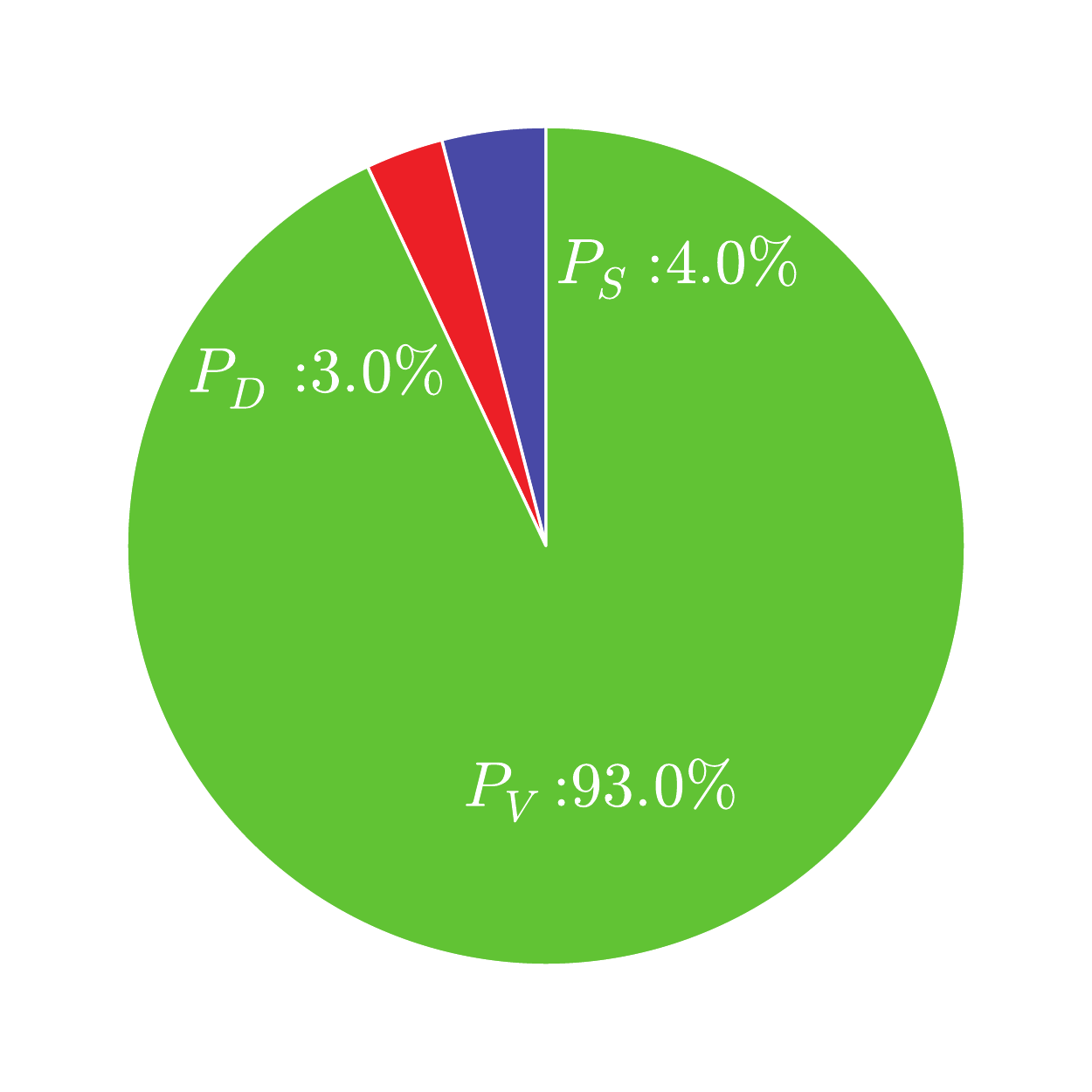}} \hspace{20mm}
\subfloat[SD-Y4O Area A]{\includegraphics[trim=8mm 8mm 8mm 8mm,clip,width=0.45\columnwidth]{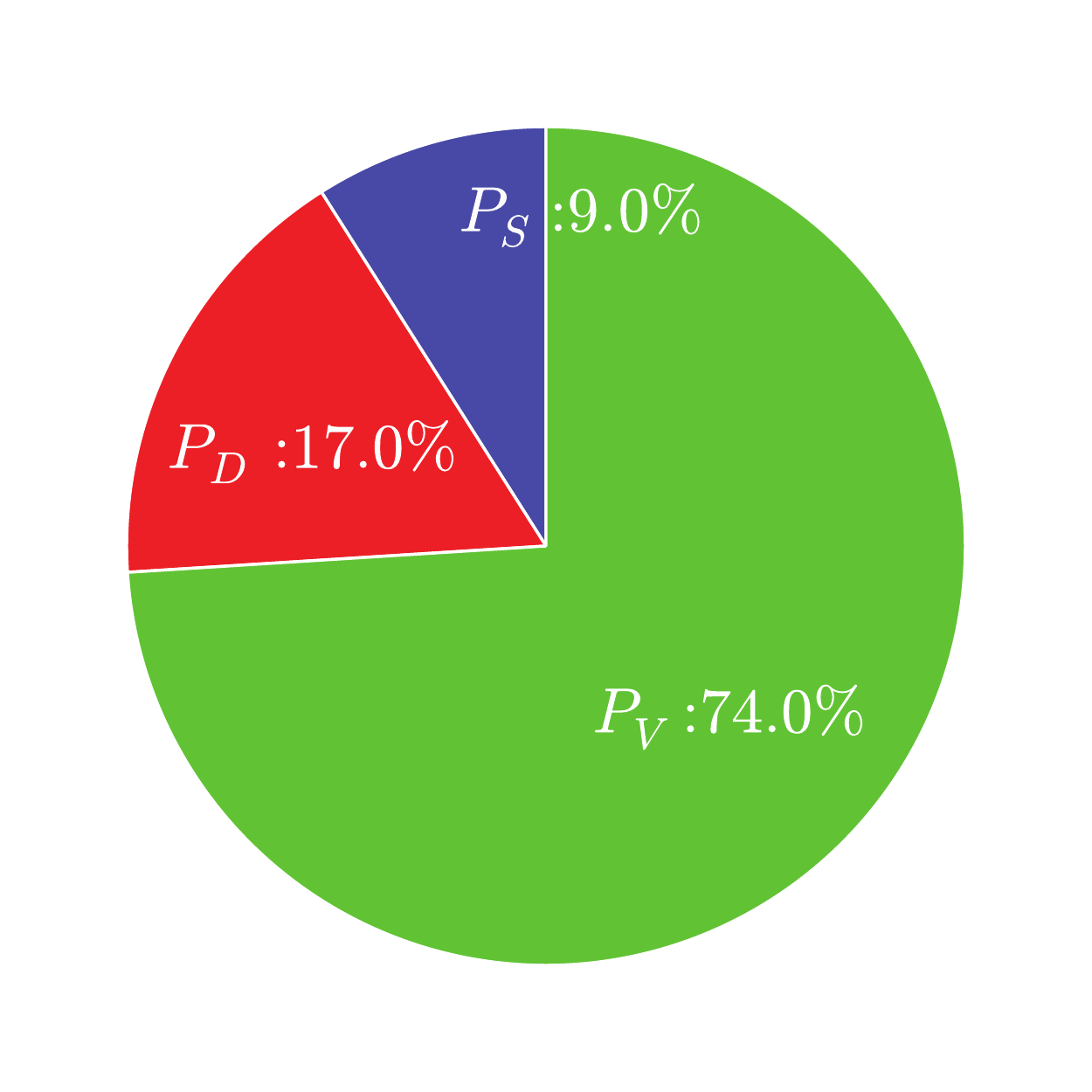}} \\
\subfloat[Y4O Area B]{\includegraphics[trim=12mm 12mm 12mm 12mm,clip,width=0.45\columnwidth]{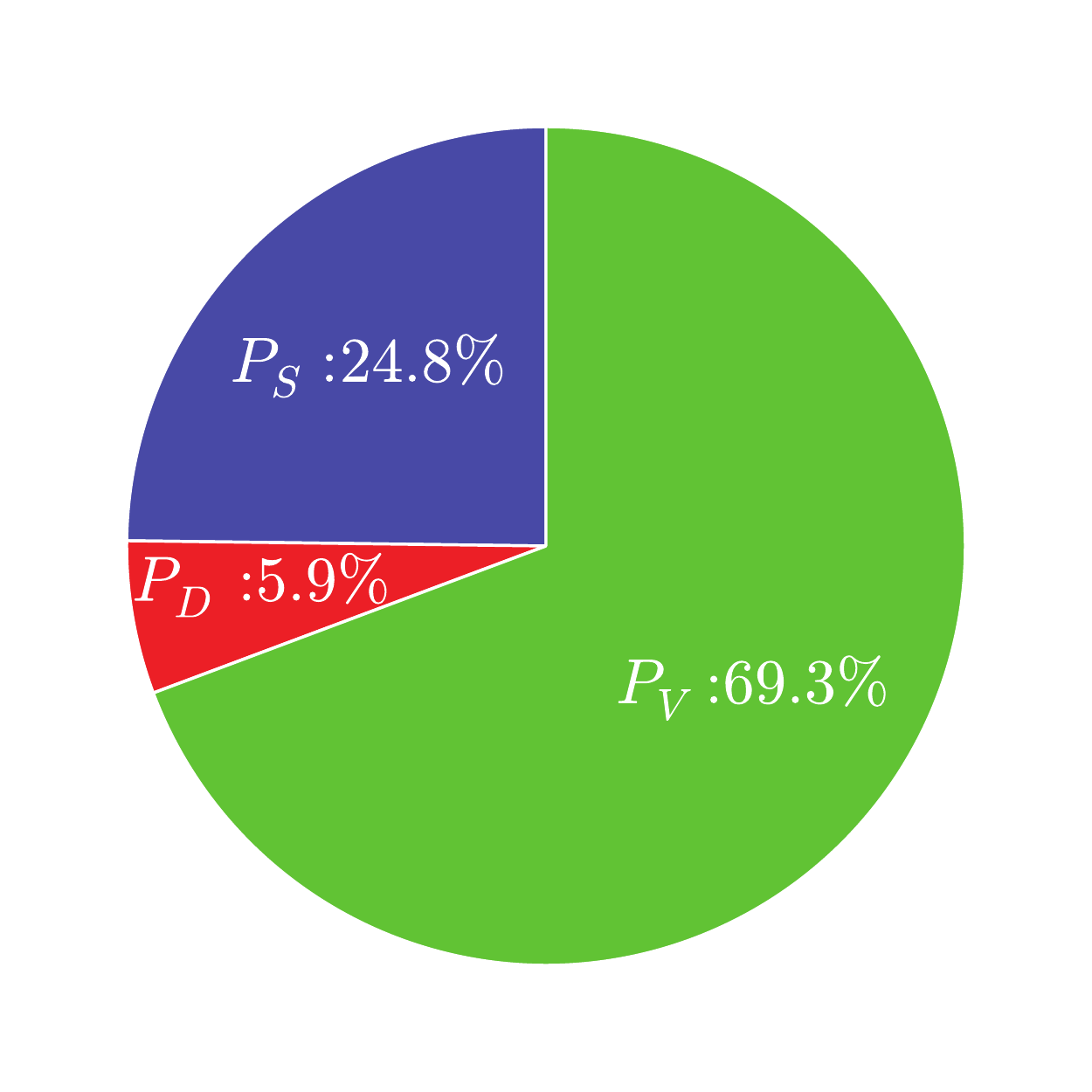}} \hspace{20mm}
\subfloat[Y4R Area B]{\includegraphics[trim=12mm 12mm 12mm 12mm,clip,width=0.45\columnwidth]{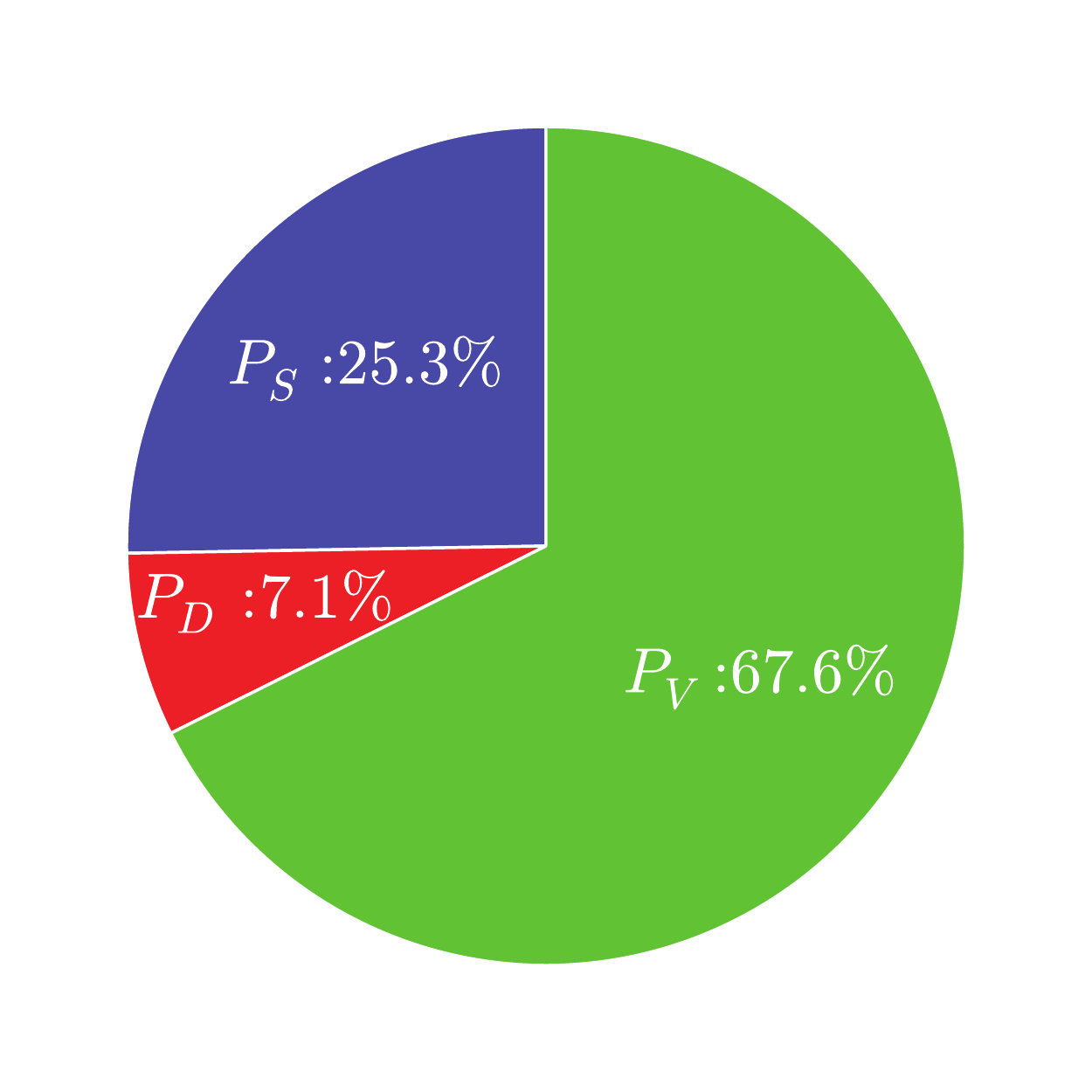}} \hspace{20mm} 
\subfloat[SD-Y4O Area B]{\includegraphics[trim=12mm 12mm 12mm 12mm,clip,width=0.45\columnwidth]{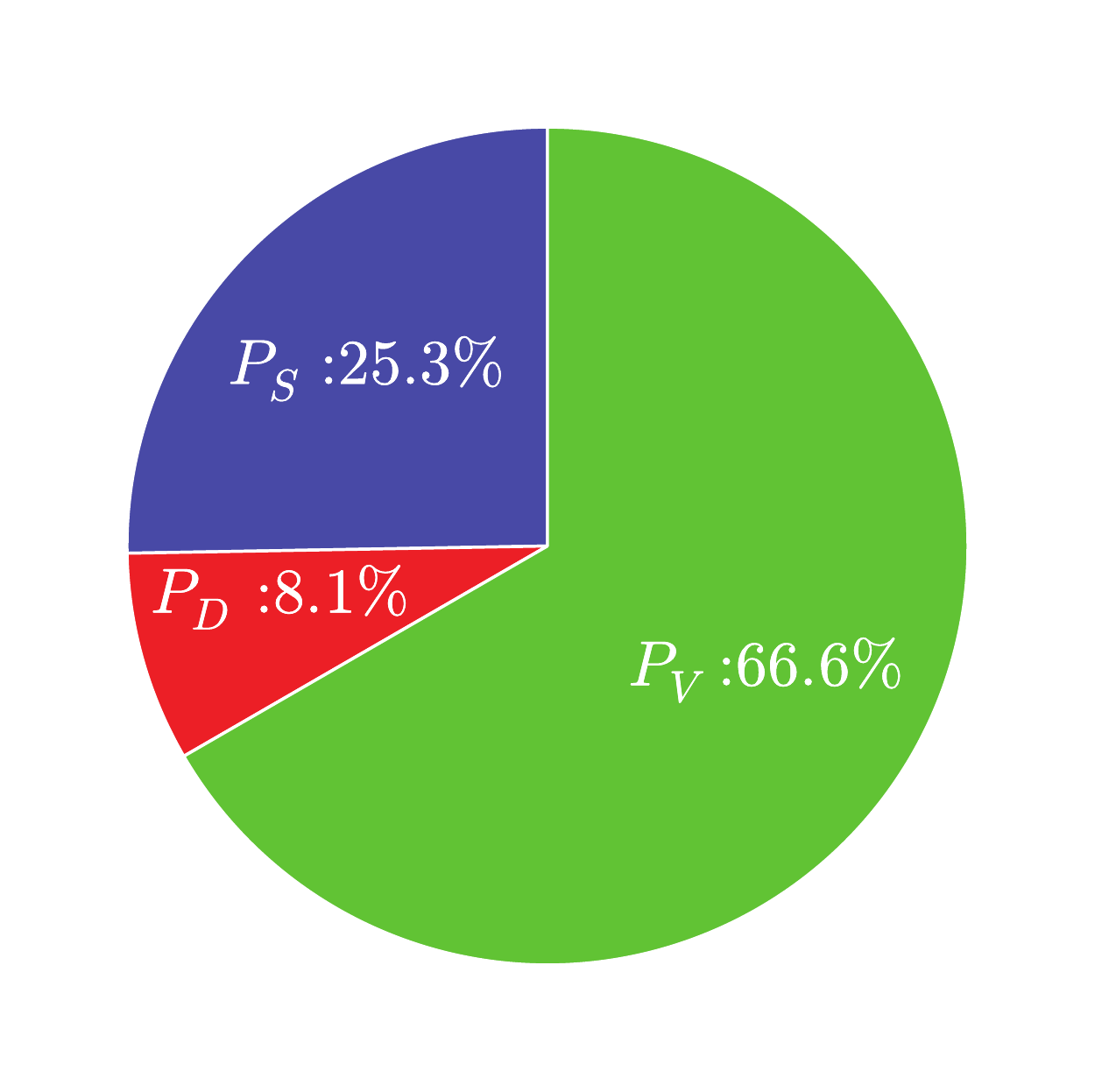}} \\
\subfloat[Y4O Area C]{\includegraphics[trim=8mm 8mm 8mm 8mm,clip,width=0.45\columnwidth]{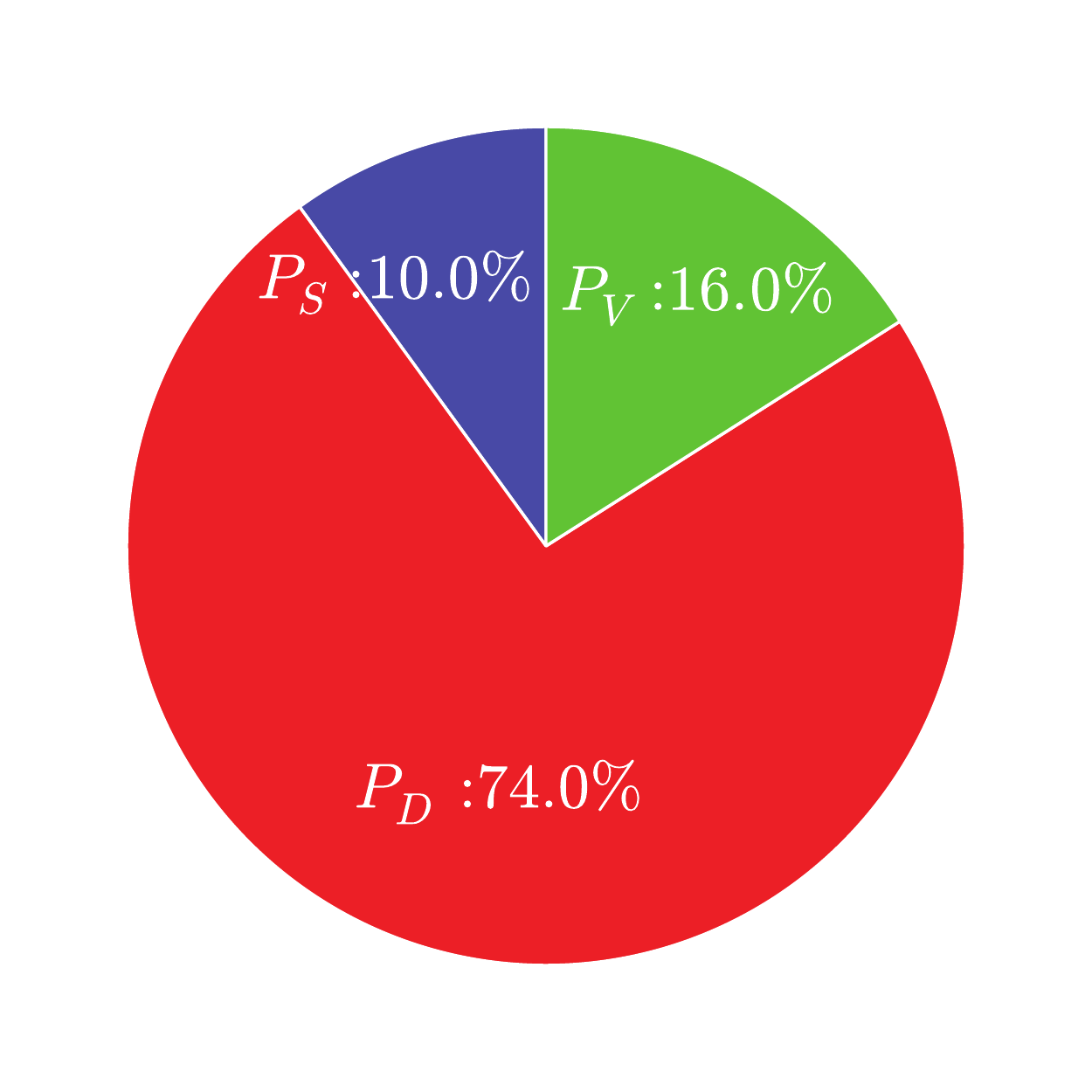}} \hspace{20mm}
\subfloat[Y4R Area C]{\includegraphics[trim=8mm 8mm 8mm 8mm,clip,width=0.45\columnwidth]{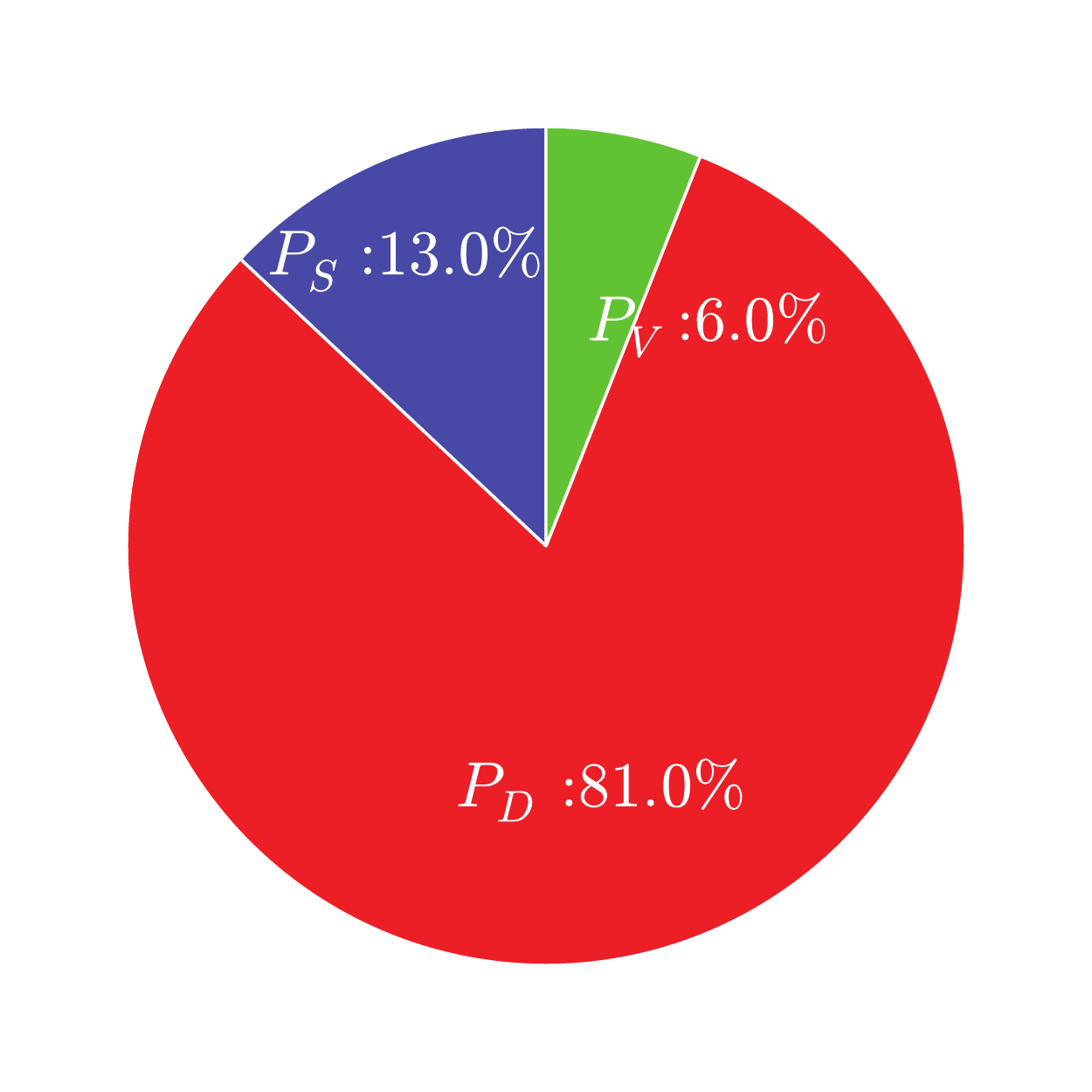}} \hspace{20mm}
\subfloat[SD-Y4O Area C]{\includegraphics[trim=8mm 8mm 8mm 8mm,clip,width=0.45\columnwidth]{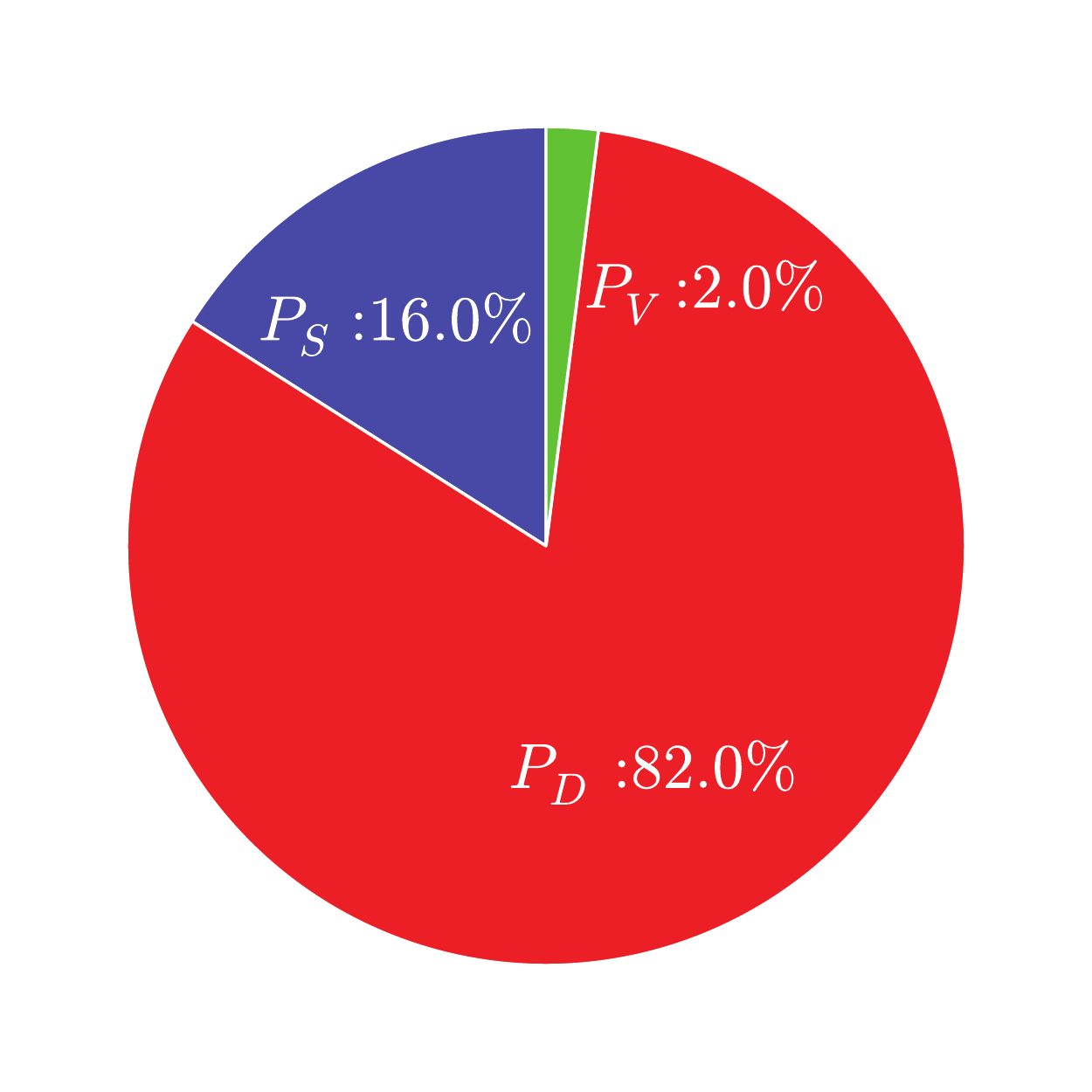}} 
\caption{Radarsat-2 C-band San-Francisco image (a)-(d)-(g)-(j): Yamaguchi four component decomposition without rotation of the coherency matrix (Y4O) for area 'A', 'B' and 'C' respectively, (b)-(e)-(h)-(k): Yamaguchi four component decomposition with rotation of the coherency matrix (Y4R) for area 'A', 'B' and 'C' respectively, (c)-(f)-(i)-(l): Y4O decomposition scattering powers modified with the Hellinger distance (SD-Y4O) for area 'A', 'B' and 'C' respectively.}
\label{fig:decom_results_1}
\end{figure*}

\begin{figure*}[!h]
\centering
\subfloat[Y4O]{\includegraphics[width=0.65\columnwidth]{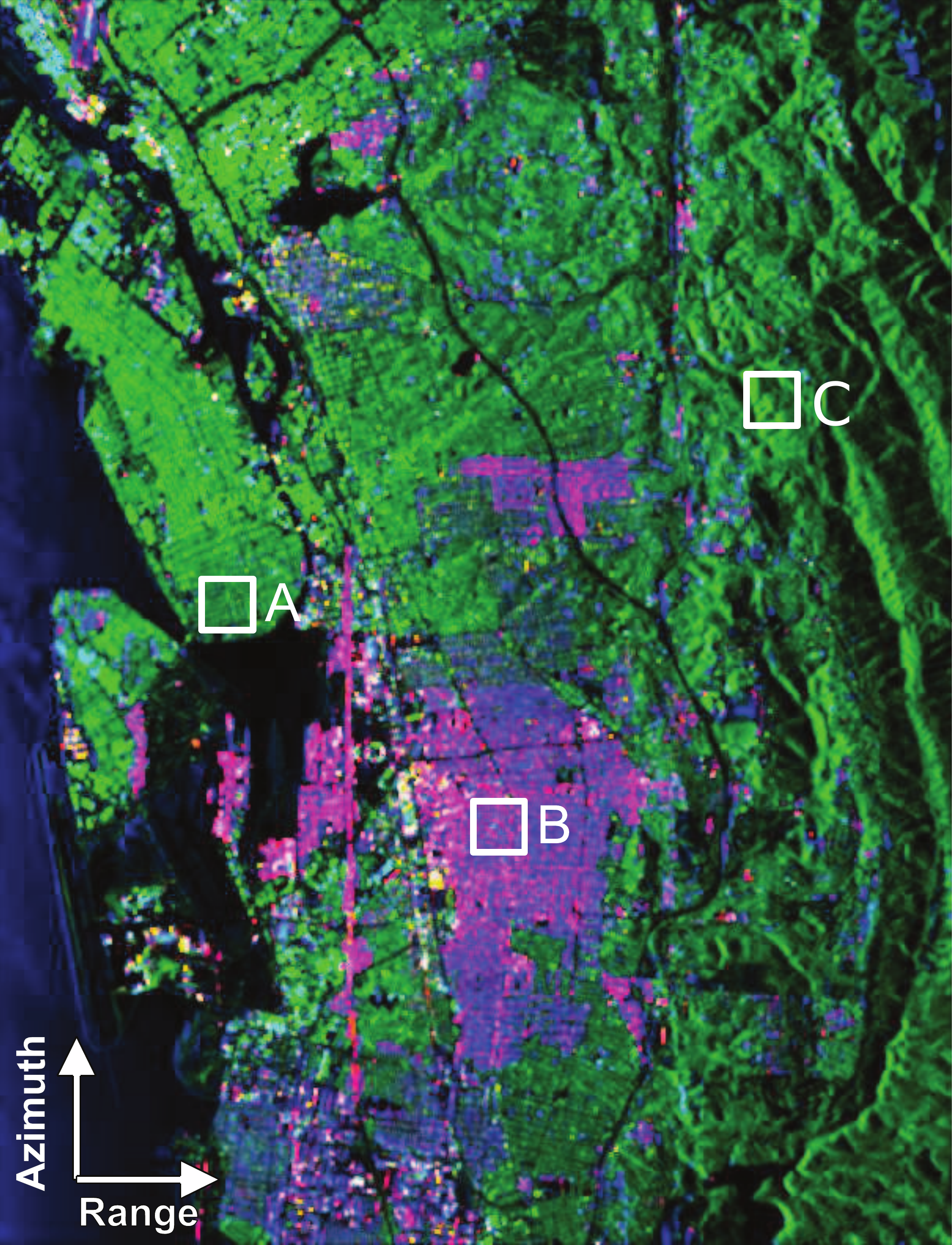}} \hspace{1mm}
\subfloat[Y4R]{\includegraphics[width=0.65\columnwidth]{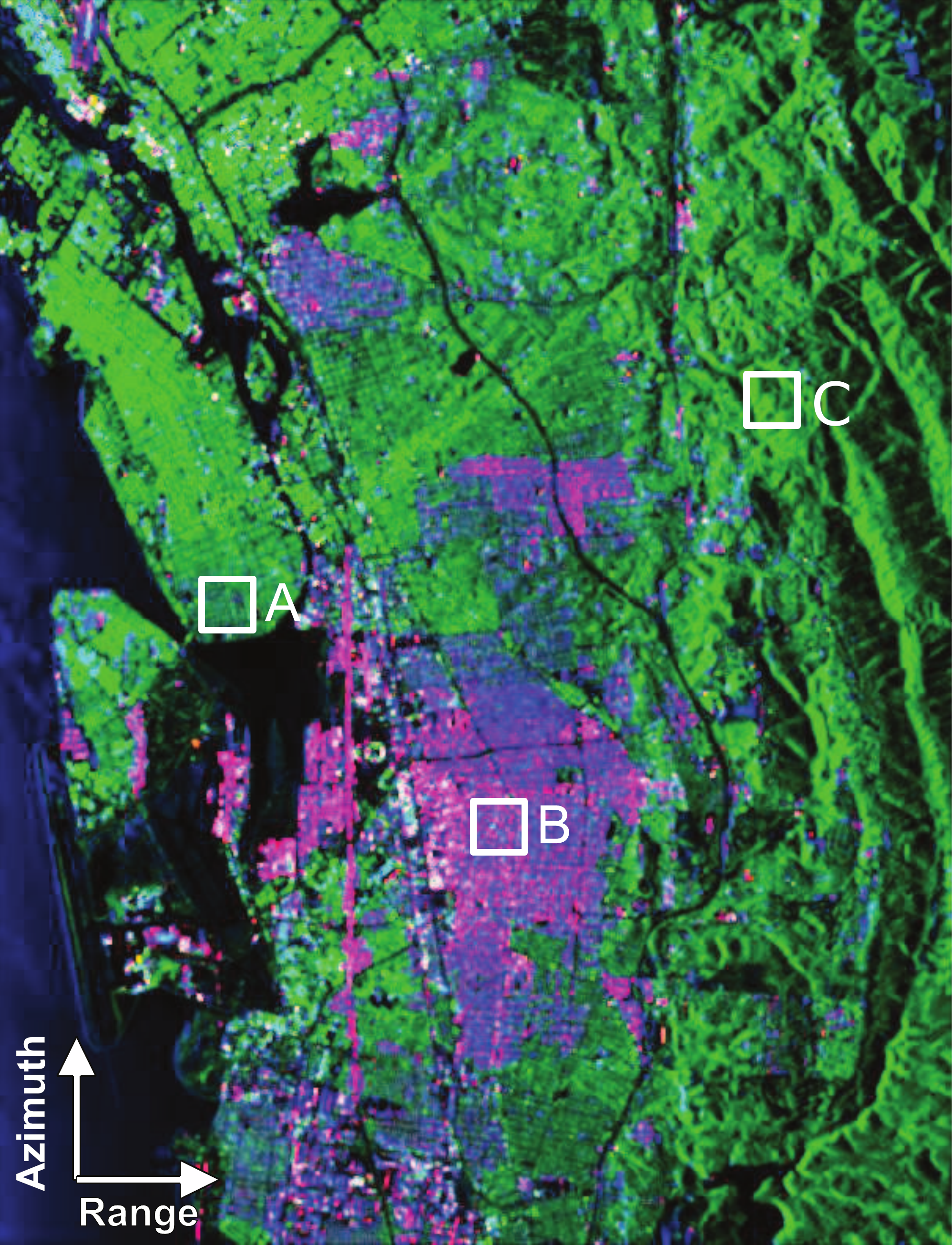}} \hspace{1mm}
\subfloat[SD-Y4O]{\includegraphics[width=0.65\columnwidth]{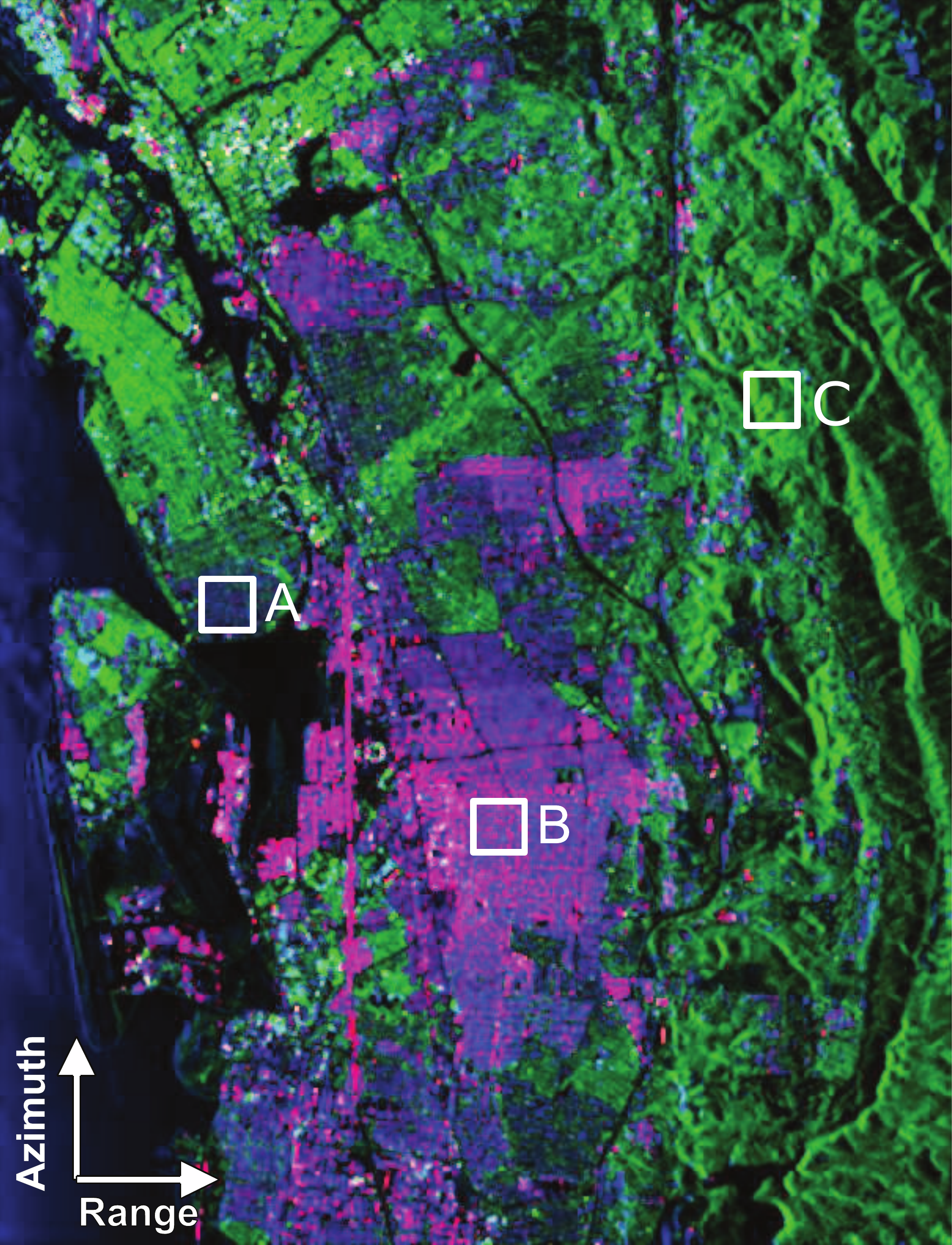}} \\
\subfloat[Y4O Area A]{\includegraphics[trim=13mm 13mm 13mm 13mm,clip,width=0.45\columnwidth]{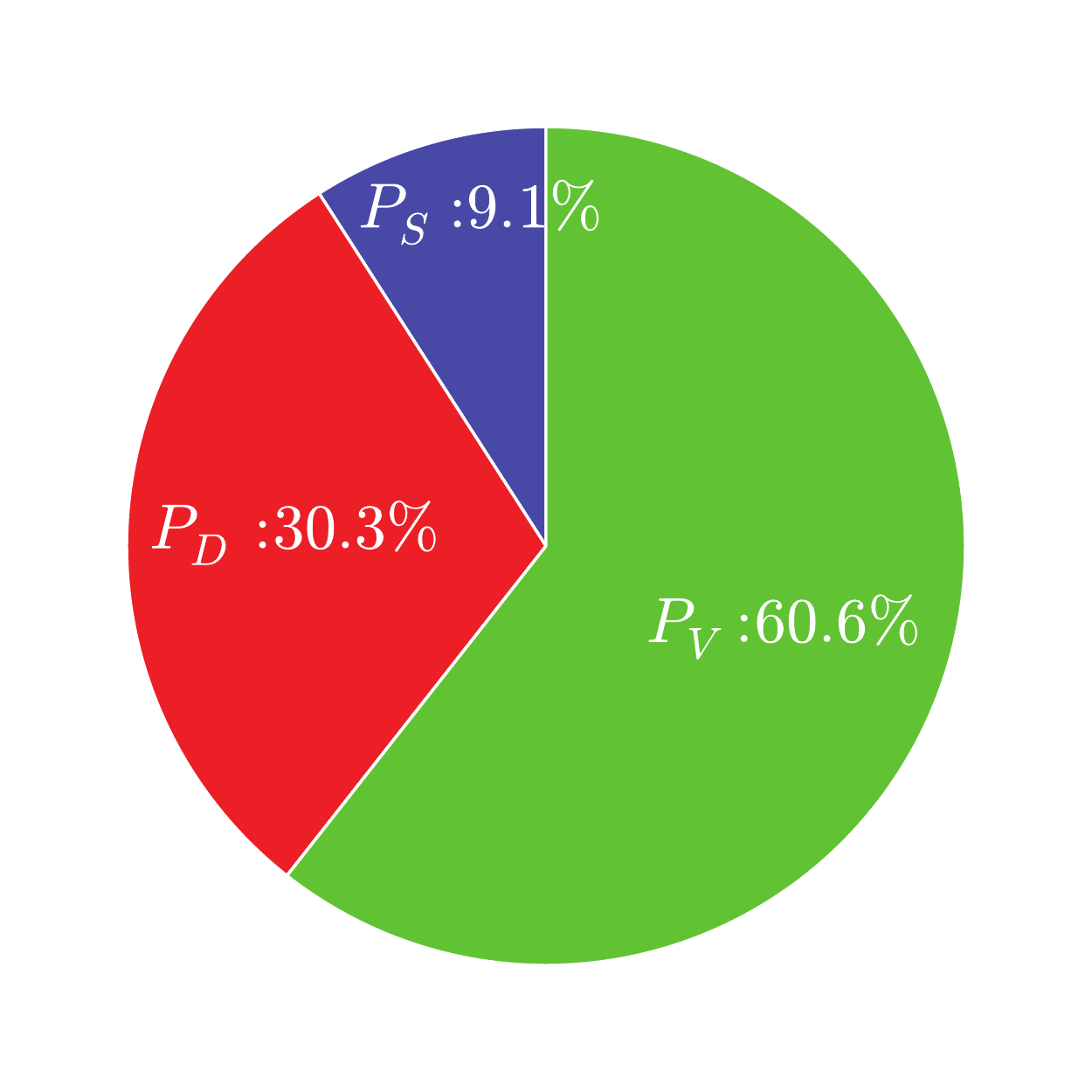}} \hspace{20mm}
\subfloat[Y4R Area A]{\includegraphics[trim=13mm 13mm 13mm 13mm,clip,width=0.45\columnwidth]{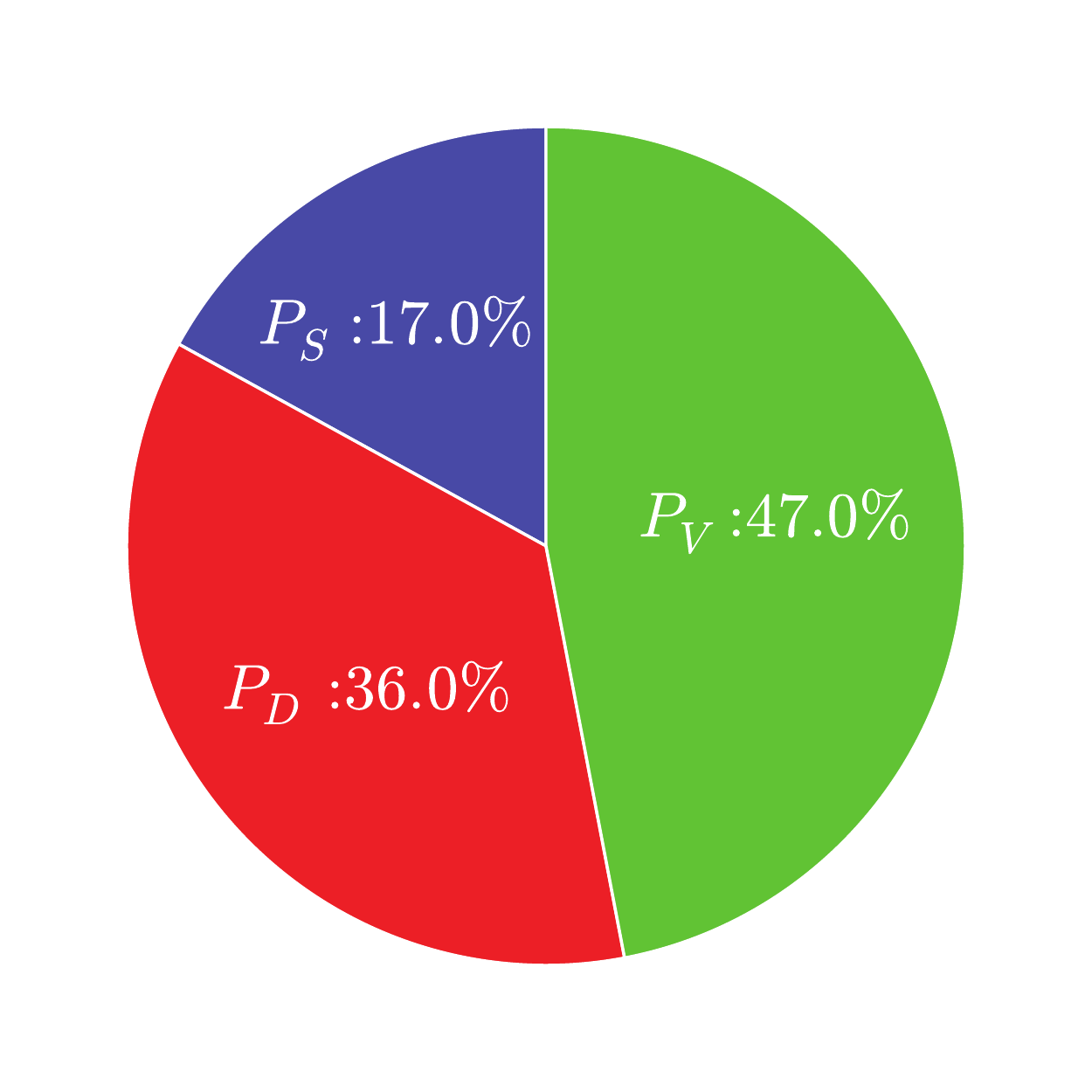}} \hspace{20mm}
\subfloat[SD-Y4O Area A]{\includegraphics[trim=13mm 13mm 13mm 13mm,clip,width=0.45\columnwidth]{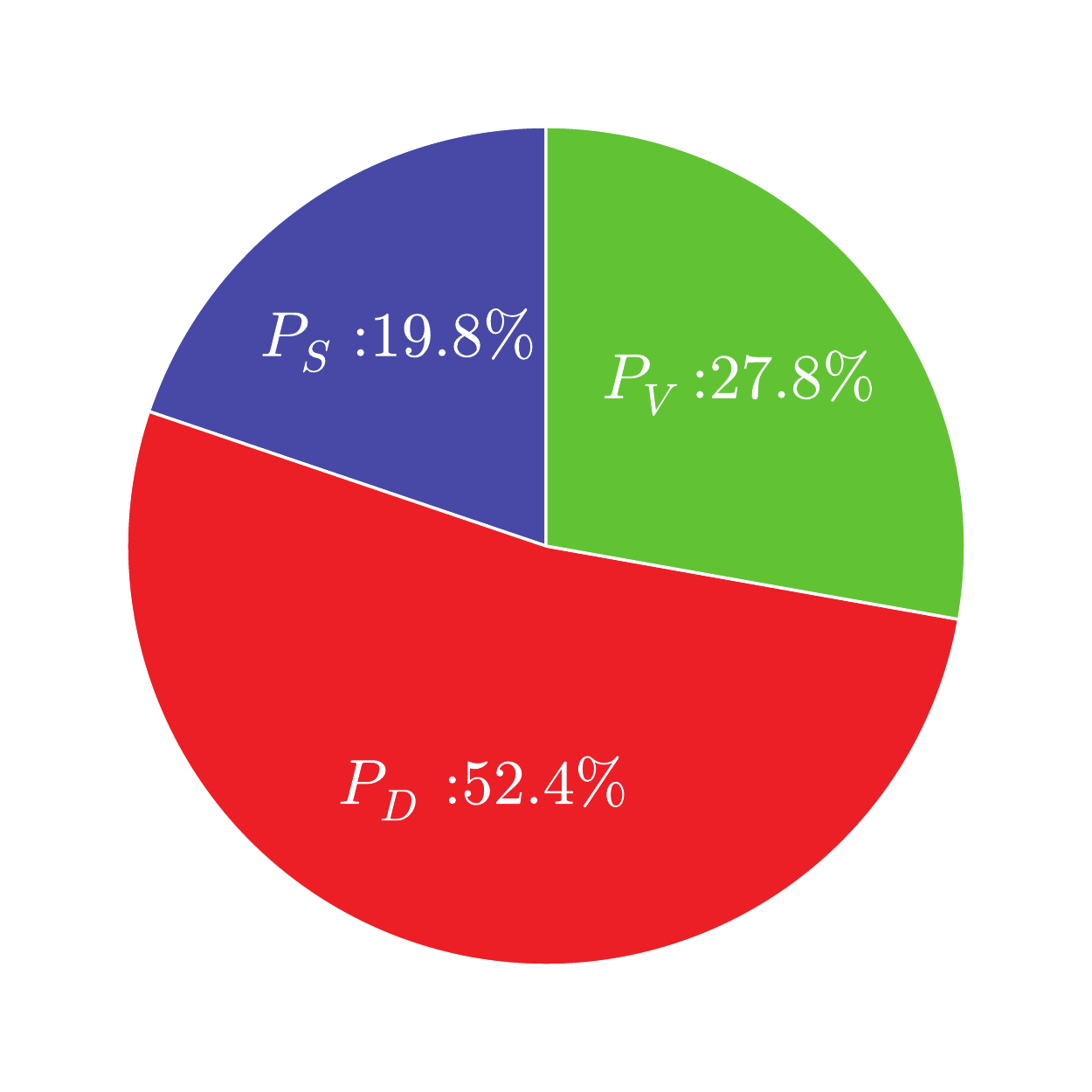}} \\
\subfloat[Y4O Area B]{\includegraphics[trim=13mm 13mm 13mm 13mm,clip,width=0.45\columnwidth]{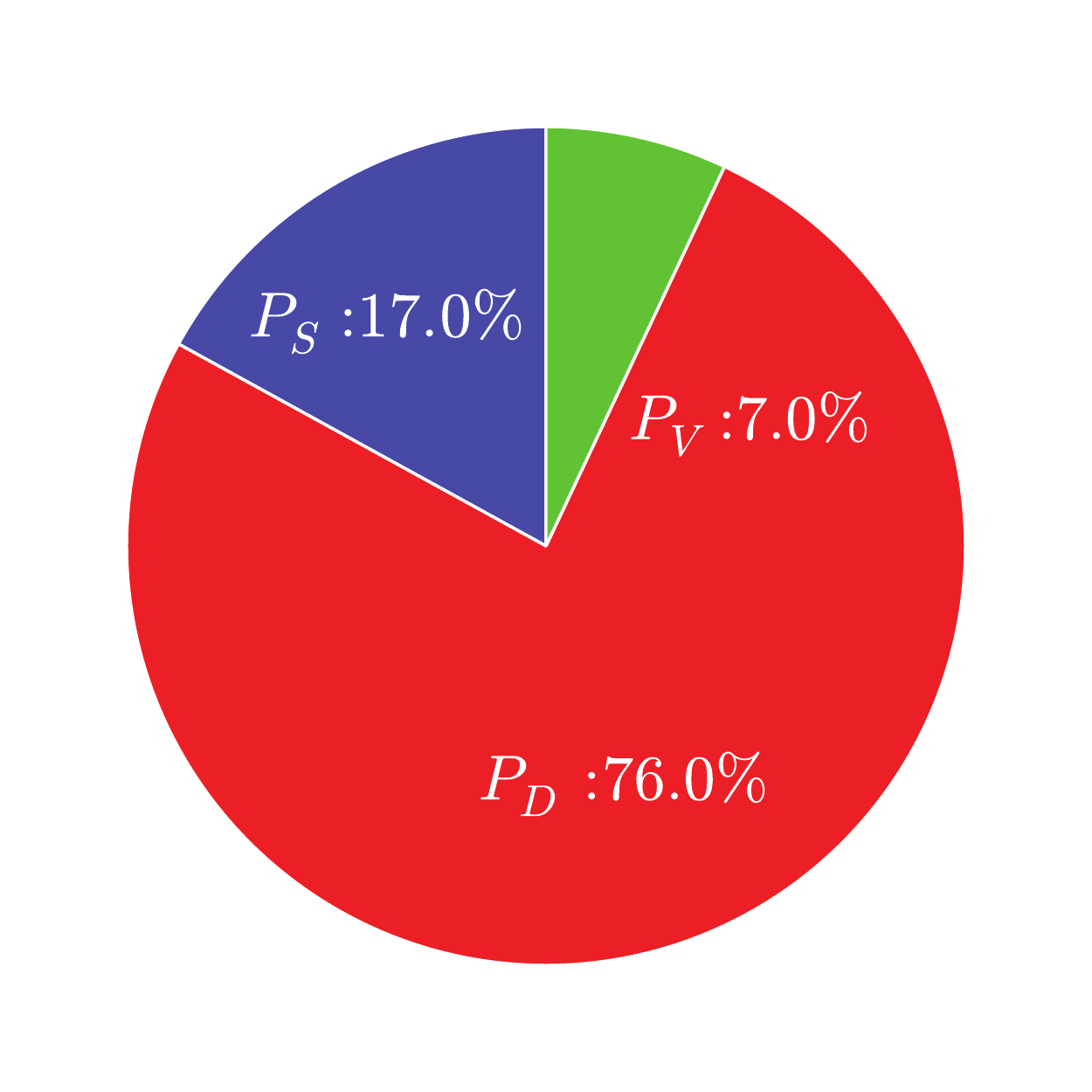}} \hspace{20mm}
\subfloat[Y4R Area B]{\includegraphics[trim=13mm 13mm 13mm 13mm,clip,width=0.45\columnwidth]{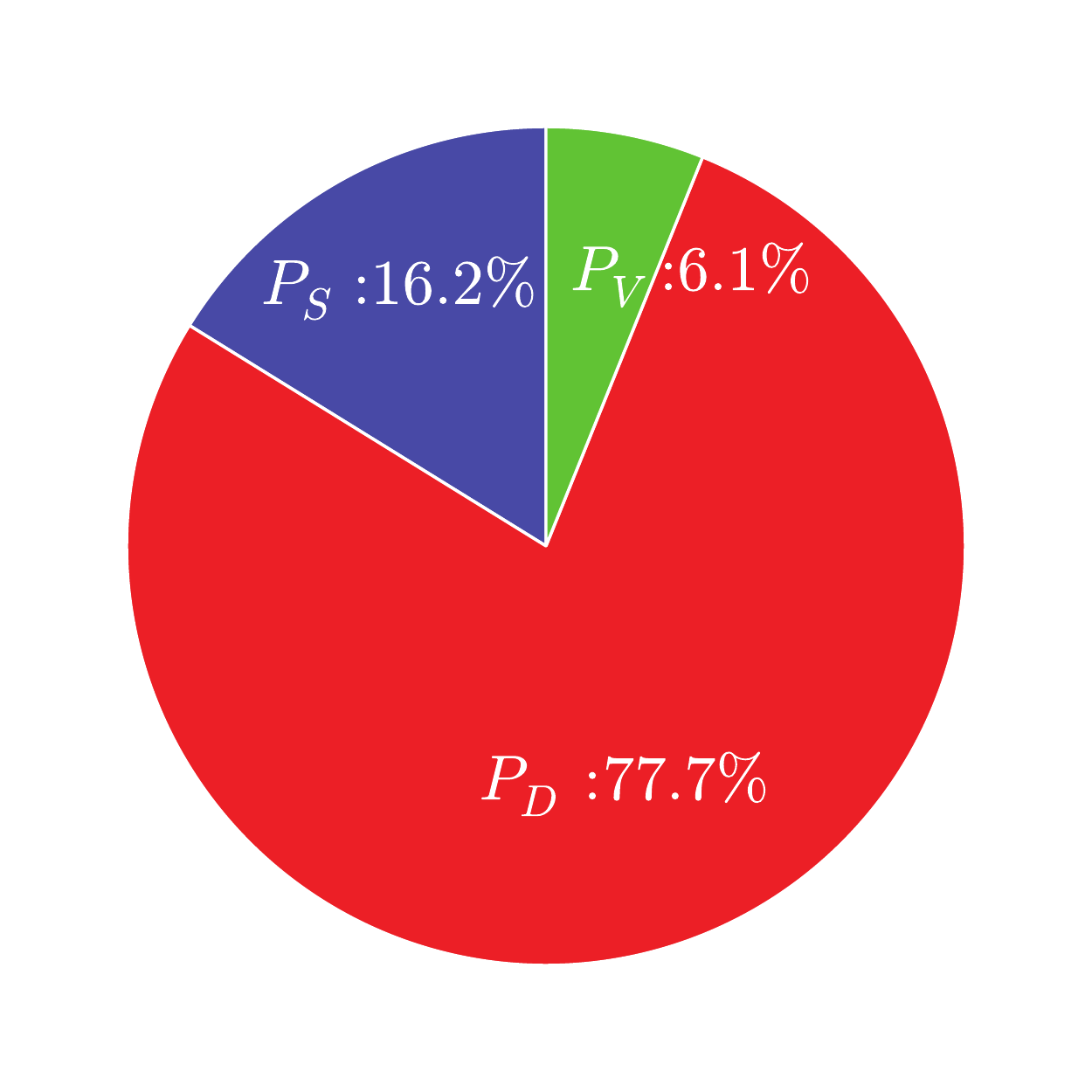}} \hspace{20mm}
\subfloat[SD-Y4O Area B]{\includegraphics[trim=13mm 13mm 13mm 13mm,clip,width=0.45\columnwidth]{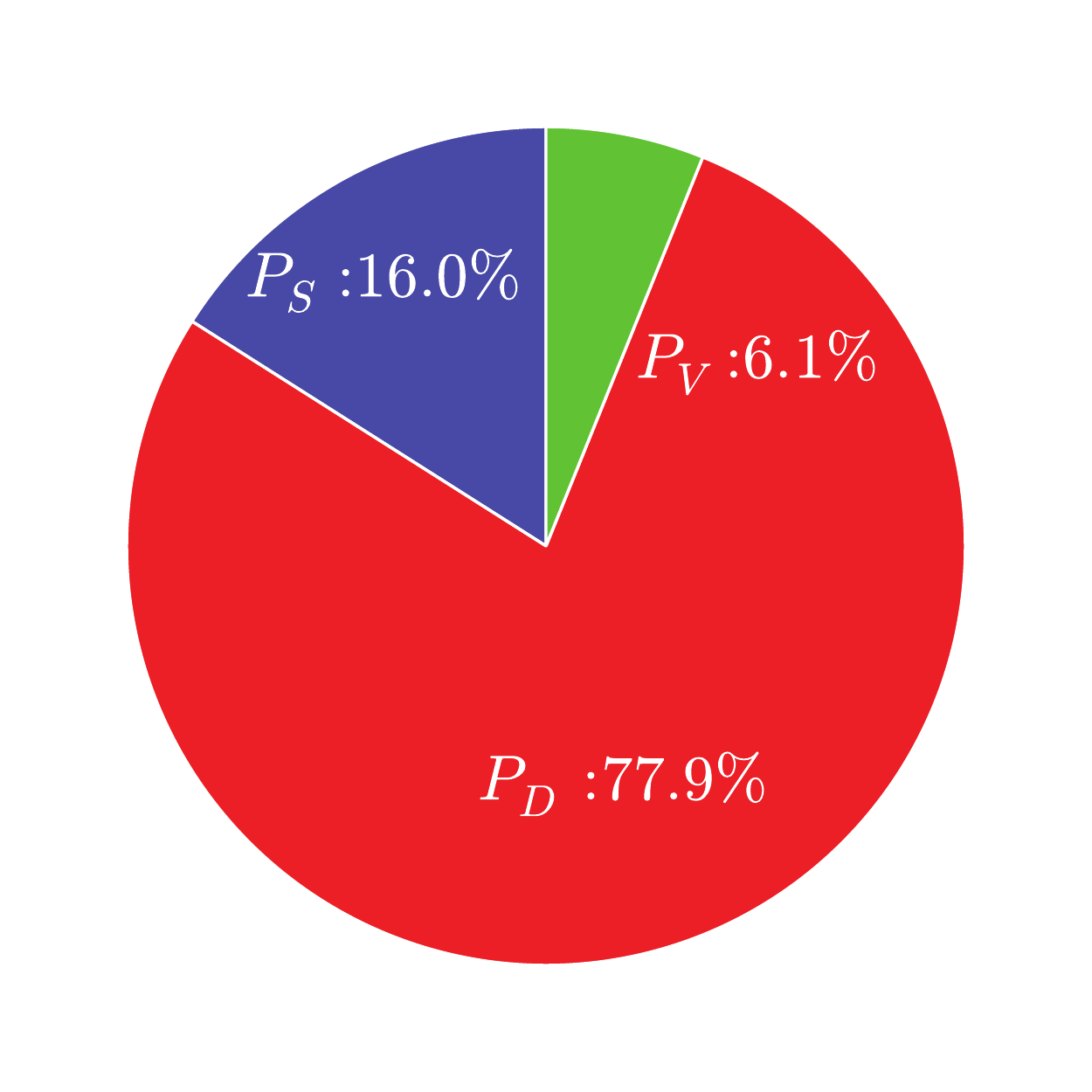}} \\
\subfloat[Y4O Area C]{\includegraphics[trim=13mm 13mm 13mm 13mm,clip,width=0.45\columnwidth]{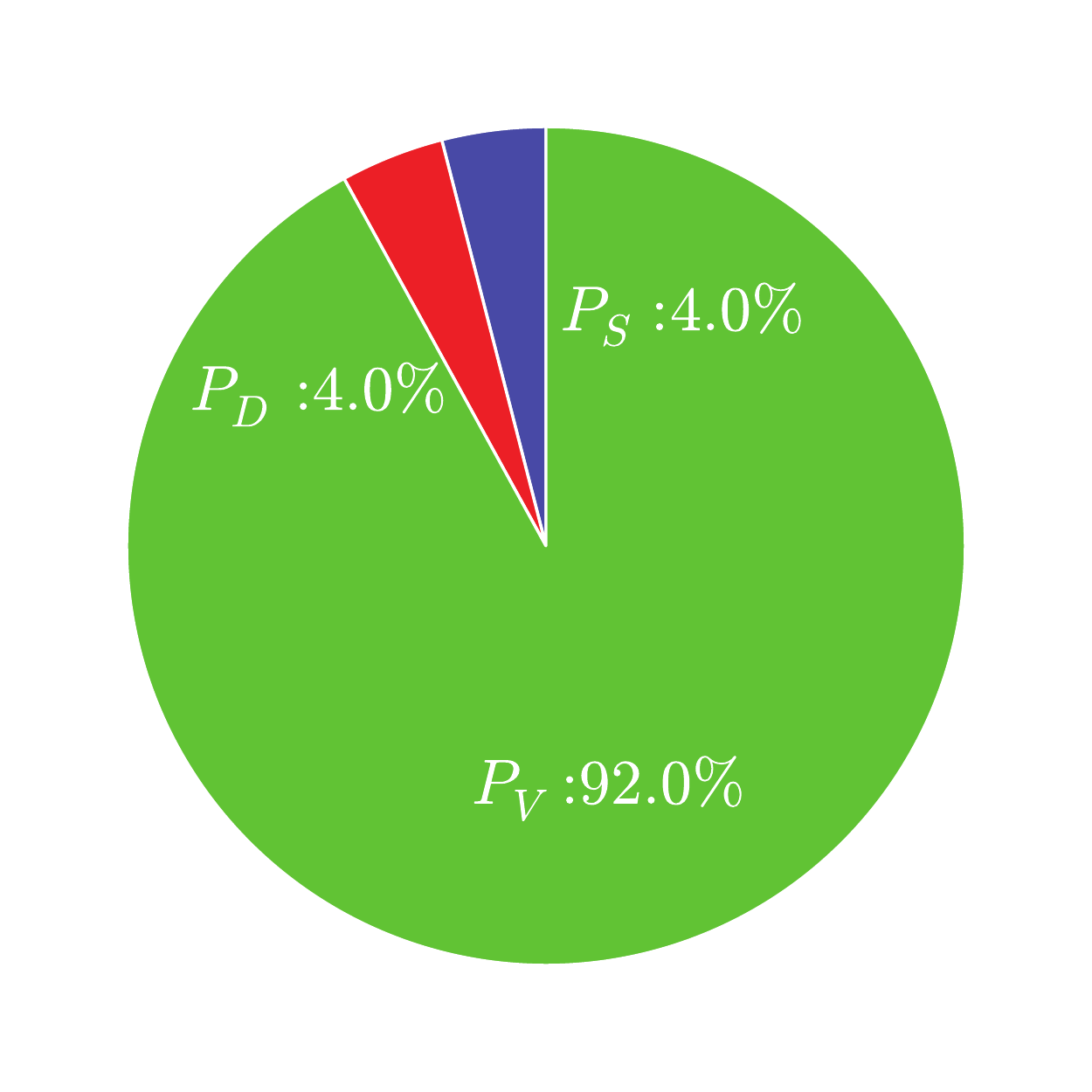}} \hspace{20mm}
\subfloat[Y4R Area C]{\includegraphics[trim=13mm 13mm 13mm 13mm,clip,width=0.45\columnwidth]{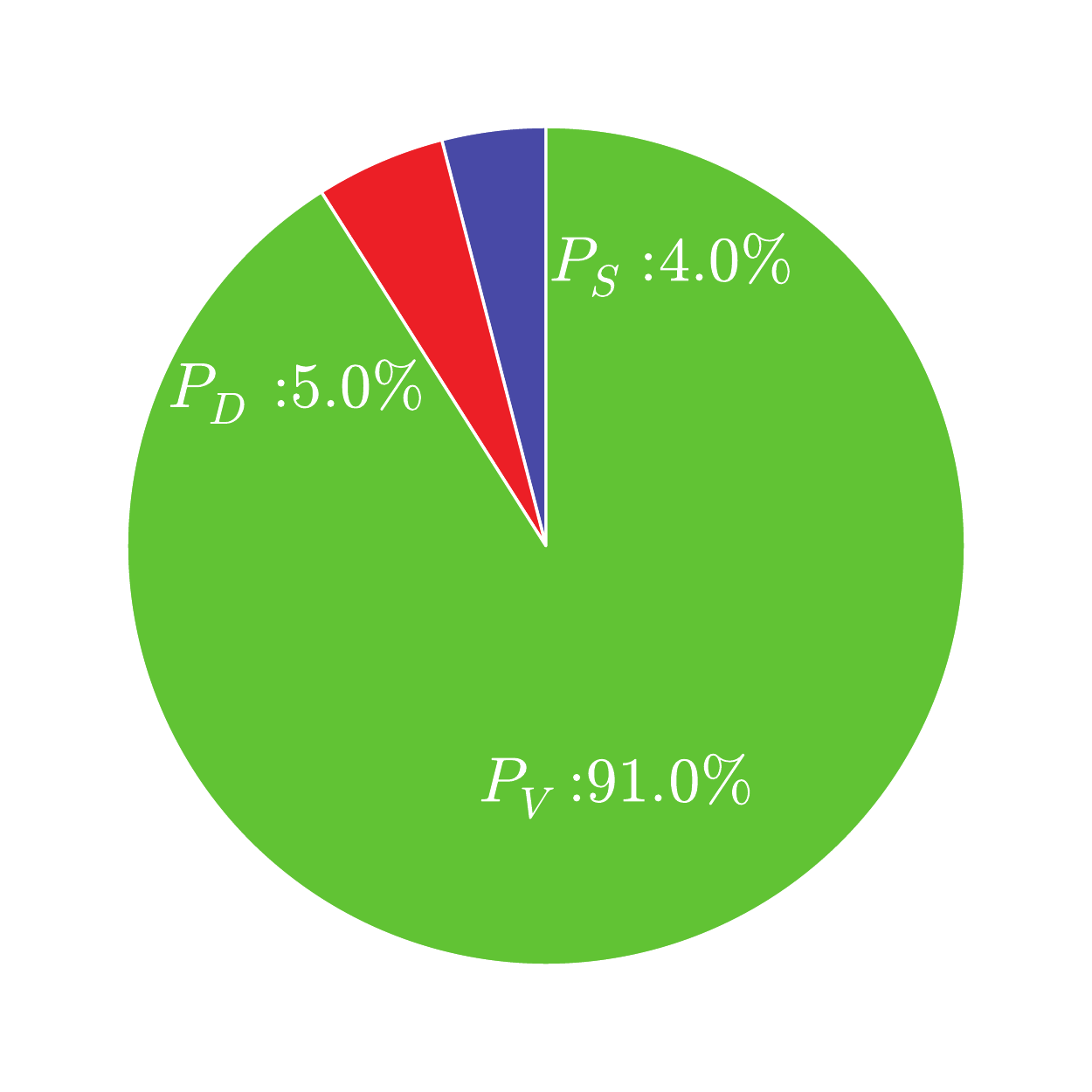}} \hspace{20mm}
\subfloat[SD-Y4O Area C]{\includegraphics[trim=13mm 13mm 13mm 13mm,clip,width=0.45\columnwidth]{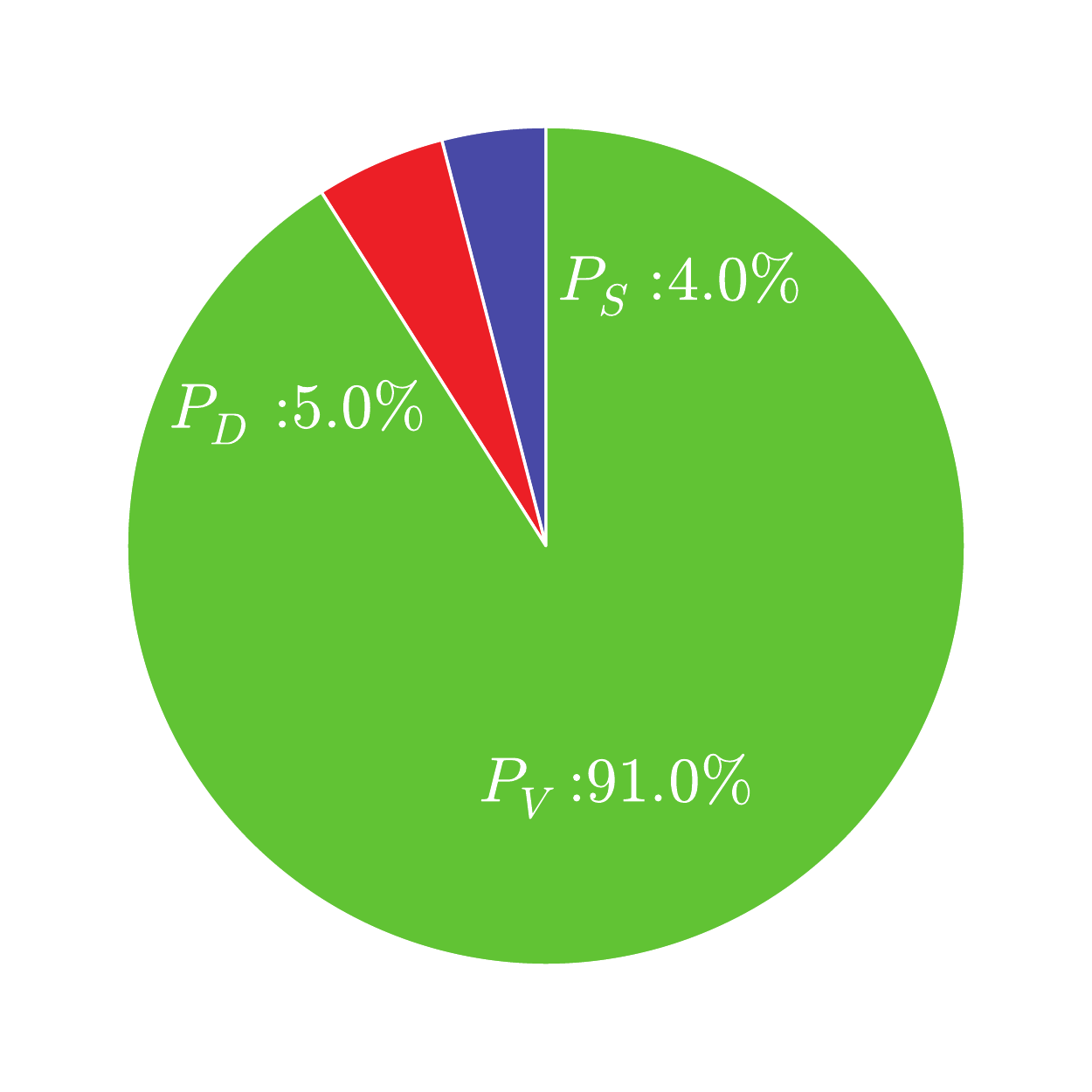}} 
\caption{UAVSAR L-band Hayward area image (a)-(d)-(g)-(j): Yamaguchi four component decomposition without rotation of the coherency matrix (Y4O) for area 'A', 'B' and 'C' respectively, (b)-(e)-(h)-(k): Yamaguchi four component decomposition with rotation of the coherency matrix (Y4R) for area 'A', 'B' and 'C' respectively, (c)-(f)-(i)-(l): Y4O decomposition scattering powers modified with the Hellinger distance (SD-Y4O) for area 'A', 'B' and 'C' respectively.}
\label{fig:decom_results_2}
\end{figure*}

\section{Results and Discussion}\label{Sec:results_discussion}
A fine-beam quad polarization (FQ9) Radarsat-2 C-band image over San Francisco with a spatial resolution of $\SI{8}{\meter}\times\SI{8}{\meter}$, and a UAVSAR L-band data with a spatial resolution of $\SI{12}{\meter}\times\SI{12}{\meter}$ (multilooked 3 times in both range and azimuth) are used in this study. 
The $P_s$, $P_d$ and $P_v$ powers are obtained from the Y4O and Y4R decomposition algorithms given in~\cite{Yamaguchi05} and~\cite{YAMAGUCHI2011}, respectively. 
The $P_s$, $P_d$ and the $P_v$ power components corresponding to the two decompositions (Y4O and Y4R) and the proposed modification (SD-Y4O) are shown in Fig.~\ref{fig:decom_results_1} for the Radarsat-2 C-band San-Francisco image. 
The Pauli RGB compositions of the Radarsat-2 and the UAVSAR images are shown in Fig.~\ref{fig:SF_UAVSAR}, 
along with high resolution optical images of patches of urban areas (both aligned and rotated with respect to the LOS), and of a forested area.

Three types of land cover classes are considered in the Radarsat-2 image: (1)~Area ``A'' is a highly oriented urban region, (2)~Area ``B'' is a region with vegetation cover, and (3)~Area ``C'' is the Golden Gate bridge. 
Over area ``A'', the double-bounce power has increased from $2\%$ for Y4O to $17\%$ for SD-Y4O, and the surface power has increased from $3\%$ for Y4O to $9\%$ for SD-Y4O, as shown in Fig.~\ref{fig:decom_results_1}(d)-(f). 
However, as expected, the powers over area ``B'' remain unchanged. 
The double-bounce power over the area ``C'' has increased from $74\%$ for Y4O to $82\%$ for SD-Y4O. A 
small change of $3\%$ for the surface power from Y4O to SD-Y4O can also be observed over this area. 
This could be attributed to the water body partly present in the area marked as ``C''.

Also three samples from the L-band UAVSAR image of the Hayward region of San Francisco are analyzed.
In this region, there are certain urban areas which are approximately $\ang{10}$--$\ang{20}$
rotated away from the LOS, as can be seen in Fig.~\ref{fig:theory_results}(a)-(c). 
A small sample, identified as ``A'', is used from this oriented urban area to compare the powers obtained from Y4O, Y4R and SD-Y4O. 
A major improvement can be seen in ``A'' with the reduction of the volume power from $60 \%$ for Y4O to $27\%$ for SD-Y4O. 
The double-bounce power has increased from $30\%$ to $52\%$ compared to Y4R, which has only increased to $36\%$, as shown in Fig.~\ref{fig:decom_results_2}(d)-(f) and in Table~\ref{table:area_A}. 
These improvements can be correctly attributed to the fact that the area under consideration is a dense urban area rotated about the LOS. 

The surface scattering power has also improved from $9.1\%$ for Y4O to $19.8\%$ for SD-Y4O. 
However, it can be observed that for the urban area ``B',' which is facing towards the LOS, the scattering powers obtained from the three methods (Y4O, Y4R and SD-Y4O) are comparable, as shown in Fig~\ref{fig:decom_results_2}(g)-(i) and in Table~\ref{table:area_B}. 
This suggests that the proposed method is useful for extracting proper scattering powers in oriented urban areas. 
In the case of non-oriented urban areas, the three methodologies provide similar results. 
Another patch, marked as ``C'' in this image, is a forested area. 
The powers obtained by the three decompositions are similar with the volume scattering power being the dominant type.

\begin{table}[hbt]
\caption{Average decomposition powers over a rotated urban area (Area 'A' marked in Fig.~\ref{fig:decom_results_2}(a)-(c))}
\centering
\begin{tabular}{c c c c}
\toprule
 Methods & $P_s$ & $P_d$ & $P_v$ \\ [0.5ex]
\midrule
Y4O & 0.05 & 0.17 & 0.33 \\ 
Y4R & 0.09 & 0.20 & 0.26\\ 
SD-Y4O & 0.11 & 0.30 & 0.14\\
\bottomrule
\end{tabular}
\label{table:area_A}
\end{table}

\begin{table}[hbt]
\caption{Average decomposition powers over an un-rotated urban area (Area 'B' marked in Fig.~\ref{fig:decom_results_2}(a)-(c))}
\centering
\begin{tabular}{c c c c}
\toprule
 Methods & $P_s$ & $P_d$ & $P_v$ \\ [0.5ex]
\midrule
Y4O & 0.37 & 1.67 & 0.16 \\ 
Y4R & 0.35 & 1.69 & 0.14\\ 
SD-Y4O & 0.38 & 1.70 & 0.10\\
\bottomrule 
\end{tabular}
\label{table:area_B}
\end{table}

\begin{table}[hbt]
\caption{Average decomposition powers over a forested area (Area ``C'' in Fig.~\ref{fig:decom_results_2}(a)-(c))}
\centering
\begin{tabular}{c c c c}
\toprule
 Methods & $P_s$ & $P_d$ & $P_v$ \\ [0.5ex]
\midrule
Y4O & 0.01 & 0.02 & 0.38 \\ 
Y4R & 0.02 & 0.02 & 0.37\\ 
SD-Y4O & 0.02 & 0.02 & 0.37\\
\bottomrule 
\end{tabular}
\label{table:area_C}
\end{table}

A major concern associated with model-based decompositions is the occurrence of negative powers in double-bounce and surface scatterings. 
Many ad-hoc techniques have been proposed which force the negative powers to be positive. 
The negative powers issue was discussed in~\cite{An10}, and in~\cite{vanZyl2011} the constraint of nonnegative eigenvalue was proposed to mitigate the problem. 
However, in this work, the double-bounce and the surface powers obtained from the original Yamaguchi four component decomposition which show negative powers are modified by the two positive quantities, $\alpha P_{v}\delta_{H}^{m}$ and $\beta P_{v}\delta_{H}^{m}$. 
Hence, it can be expected that the number of pixels with small negative powers in double-bounce and surface scattering can be made positive with the addition of the two positive quantities, but the number of pixels with large negative powers may still remain negative. 
A comparison of the negative powers for the two datasets is shown in Table~\ref{table:negative_powers}. 
It can be seen that the percentage of pixels with negative powers have decreased from Y4O to SD-Y4O for both the datasets and is comparable to Y4R. 

\begin{table}[hbt]
\caption{Percentage of pixels with negative powers}
\centering
\begin{tabular}{c c r }
\toprule
Dataset & Method & \multirow{2}{2cm}{$\%$ of Pixels with Negative Powers} \\ [1em]
\midrule
UAVSAR & Y4O & 17 \\ 
						& Y4R & 14 \\ 
						& SD-Y4O & 13 \\ \cmidrule(rl){1-3}
Radarsat-2	& Y4O & 8 \\ 
					& Y4R & 6 \\ 
					& SD-Y4O & 6 \\
\bottomrule 
\end{tabular}
\label{table:negative_powers}
\end{table}

\section{Conclusions}

The over-estimation of the volume power and, consequently, the underestimation of the double bounce and the surface powers in the Y4O decomposition in rotated urban areas is of major concern. 
In order to alleviate this issue, the Y4R decomposition was proposed. 
Several other decomposition techniques have been recently proposed which address this issue by using different scattering models. In this work we propose a stochastic distance based measure to modify the powers estimated from the Y4O decomposition for oriented targets. 

We estimate the orientation angle of the target using a criteria based on the Hellinger distance between the $T_{33}$ and $T_{22}$ elements of the coherence matrix $\mathbf{[T]}$. 
Using this stochastic distance, we have proposed a method that systematically modifies the $P_s$, $P_d$ and the $P_v$ powers to obtain better estimates. Thus, the physical property of the target, i.e., the orientation angle, is used to make modifications to the powers. The results obtained by the proposed method are encouraging.  

An L-band UAVSAR dataset over Hayward is used in this work due to the presence of rotated urban areas. 
The dataset is decomposed with the Y4R, Y4O and the proposed method. It can be seen that, on oriented urban areas, there is an increase in the $P_{s}$ power by $11\%$ and an increase in the $P_{d}$ power by $22\%$ from Y4O to SD-Y4O with a corresponding reduction of the volume power. A similar analysis is also performed on a Radatsat 2 C-band San Francisco dataset. An increase of $15\%$ in the $P_d$ power is observed for the proposed method. However, for the area with forest cover, the three powers from all the decompositions are almost identical. The authors believe that due to its simplicity, the proposed method can be easily used to modify the Y4O decomposition output, which is among the most popular model-based decomposition techniques in SAR polarimetry.  

\section*{Acknowledgement}
The authors would like to thank the anonymous reviewers for their constructive comments which
have been very useful in improving the technical quality of the manuscript.
ACF acknowledges support from CNPq, Capes and Fapeal.

\bibliographystyle{IEEEtran}
\bibliography{mybibfile}

\begin{IEEEbiography}[{\includegraphics[width=1in]{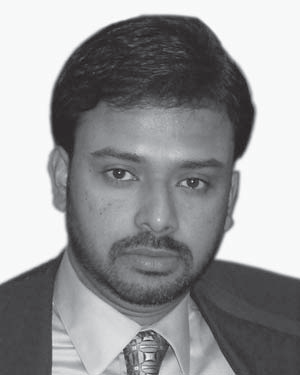}}]{Avik Bhattacharya} (M'08)
received the M.Sc. degree in Mathematics from the Indian Institute of Technology, Kharagpur, India, and the Ph.D. degree from the T{\'e}l{\'e}com ParisTech, Paris, France, and the Ariana Research Group, Institut National de Recherche en Informatique et en Automatique (INRIA), Sophia Antipolis, Nice, France.
He is currently an Assistant Professor at the Centre of Studies In Resources Engineering, Indian Institute of Technology Bombay (IITB). Prior to joining IITB, he was a Canadian government research fellow at the Canadian Centre for Remote Sensing (CCRS) in Ottawa, Canada. He has received the Natural Sciences and Engineering Research Council of Canada (NSERC) visiting scientist fellowship at the Canadian national laboratories from 2008 to 2011. His current research includes SAR polarimetric application to cryosphere and planetary exploration, statistical analysis of polarimetric SAR images, machine learning and pattern recognition.
\end{IEEEbiography}

\begin{IEEEbiography}[{\includegraphics[width=1in]{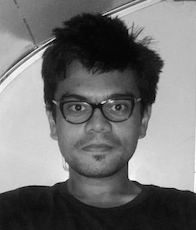}}]{Arnab Muhuri}received his
B.E degree in Electrical Engineering in 2009. He also received his Masters
in Technology (M.Tech) degree from Indian Institute of Technology Bombay, India, where he worked on statistical modelling of satellite radar images. From 2012 to
2013, he was with the Microwave Remote Sensing Laboratory, CSRE at Indian
Institute of Technology Bombay, India, as a Research Fellow, where he
worked in the field of Planetary Remote Sensing. He is currently a
PhD student at the Microwave Remote Sensing Laboratory, CSRE at
Indian Institute of Technology Bombay, India, working in
the field of applications of satellite images for monitoring cryosphere.
His research interests are in the field of microwave remote sensing and image processing. 
\end{IEEEbiography}

\begin{IEEEbiography}[{\includegraphics[width=1in]{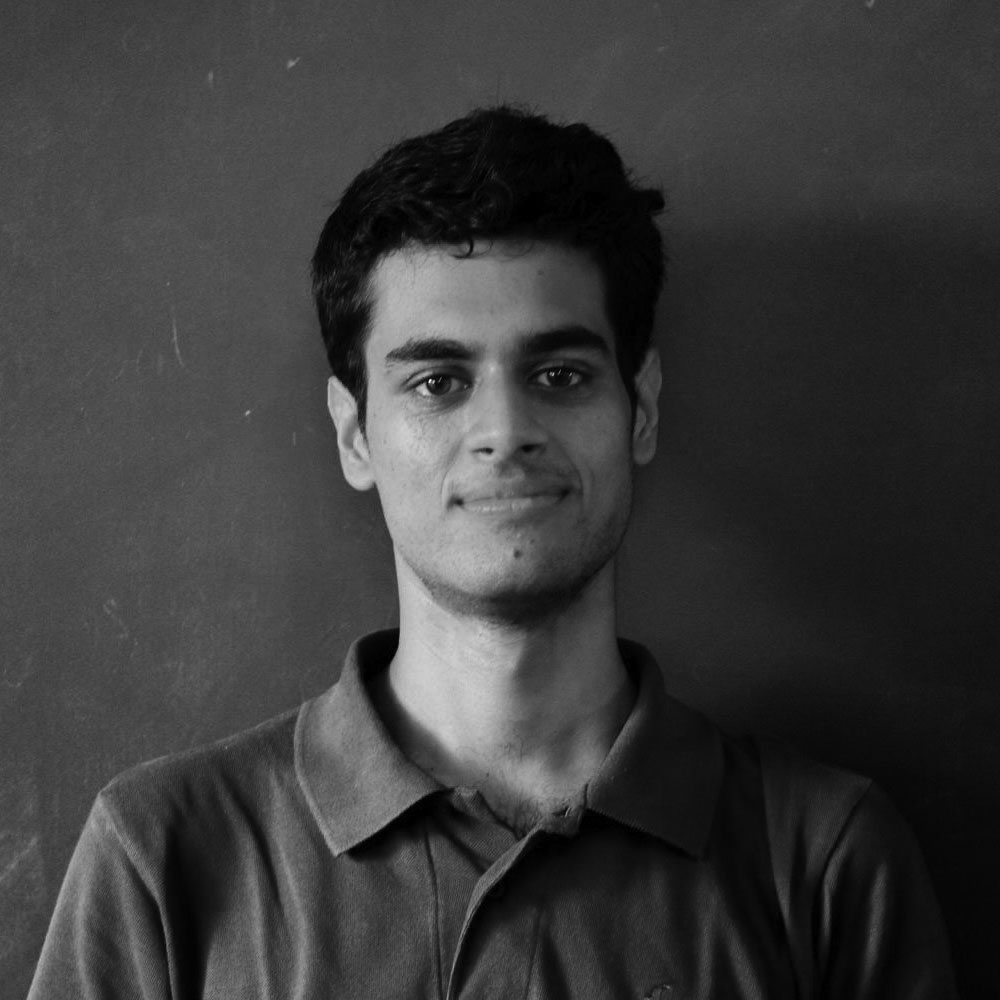}}]{Shaunak De}
received his Bachelor of Engineering in Electronics Engineering
from University of Mumbai in 1990, graduating as a gold medalist. He is
currently pursuing his PhD from the Centre of Studies in Resources
Engineering (CSRE), Indian Institute of Technology Bombay, India. His
research interests include Polarimetric SAR, Machine Learning and
Information Theory.
\end{IEEEbiography}

\begin{IEEEbiography}[{\includegraphics[width=1in]{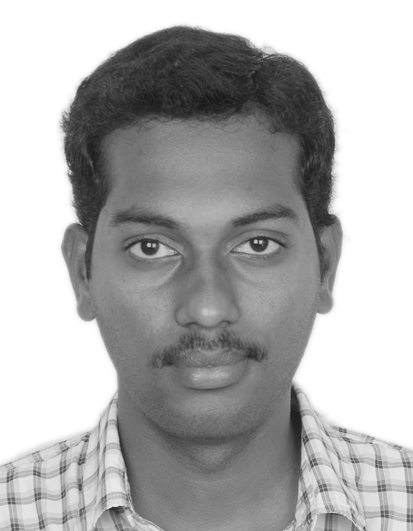}}]{Surendar Manickam} 
received the M.Sc degree in Physics from the Periyar  University and M.Tech degree  in Remote Sensing from the Anna university , Tamilnadu , India in 2008 and 2011, respectively. He is currently pursuing his PhD at the Centre of Studies in Resources Engineering (CSRE), Indian Institute of Technology Bombay, India. His current research interests are in SAR polarimetry application in cryosphere.
\end{IEEEbiography}

\begin{IEEEbiography}[{\includegraphics[width=1in]{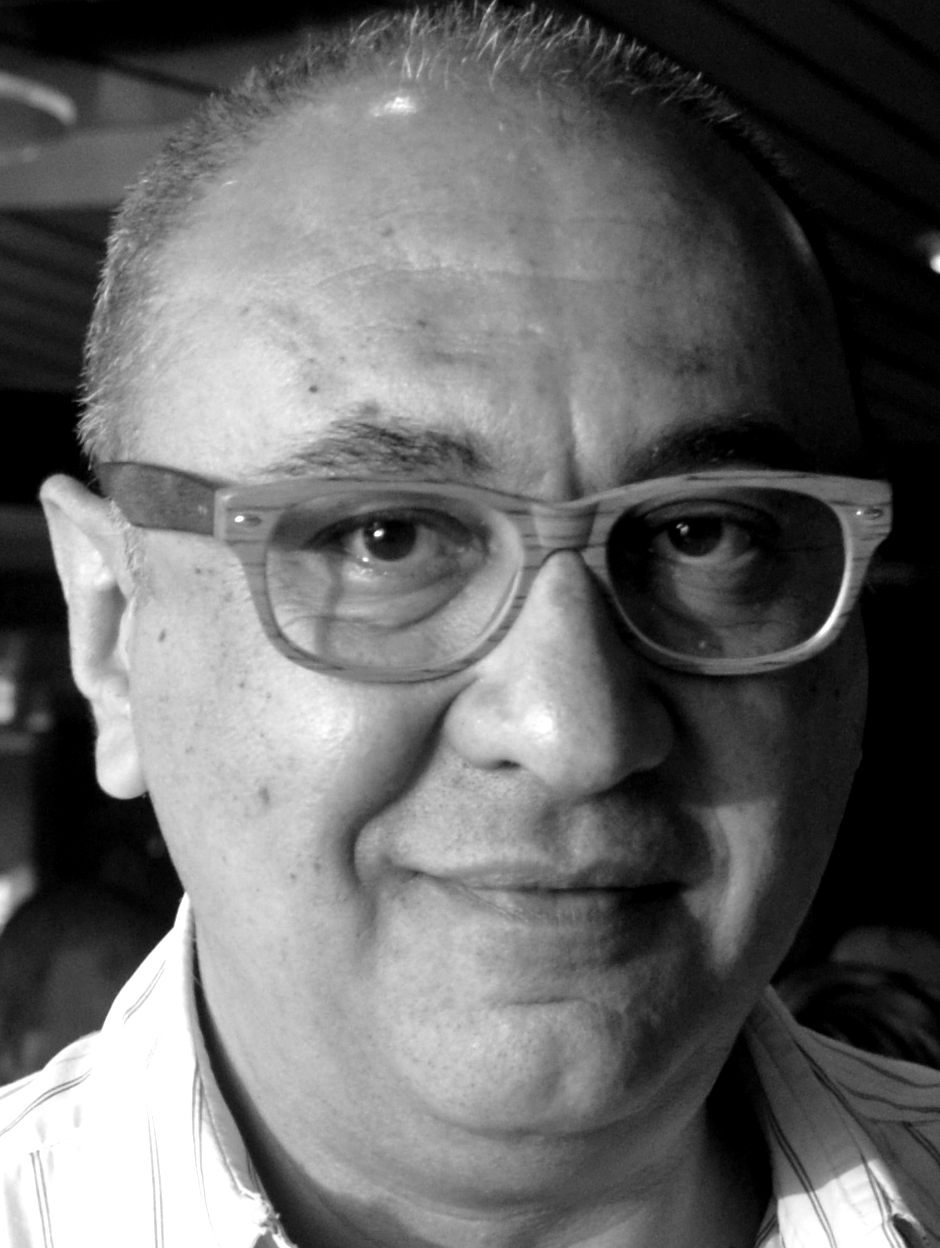}}]{Alejandro C.\ Frery} (S'92--SM'03)
received the B.Sc. degree in Electronic and Electrical Engineering from the Universidad de Mendoza, Argentina.
His M.Sc. degree was in Applied Mathematics (Statistics) from the Instituto de Matem\'atica Pura e Aplicada (IMPA, Rio de Janeiro) and his Ph.D. degree was in Applied Computing from the Instituto Nacional de Pesquisas Espaciais (INPE, S\~ao Jos\'e dos Campos, Brazil).
He is currently the leader of LaCCAN -- \textit{Laborat\'orio de Computa\c c\~ao Cient\'ifica e An\'alise Num\'erica}, Universidade Federal de Alagoas, Macei\'o, Brazil.
His research interests are statistical computing and stochastic modelling.
\end{IEEEbiography}

\end{document}